\documentclass[sigconf]{acmart}
\usepackage{graphicx}
\usepackage{subcaption}
\usepackage[utf8]{inputenc}
\usepackage{graphicx}
\usepackage{xcolor}
\usepackage{ulem}
\usepackage{caption}
\AtBeginDocument{%
  }

\copyrightyear{2026}
\acmYear{2026}
\setcopyright{cc}
\setcctype{by-nc-nd}
\acmConference[CHI '26]{Proceedings of the 2026 CHI Conference on Human Factors in Computing Systems}{April 13--17, 2026}{Barcelona, Spain}
\acmBooktitle{Proceedings of the 2026 CHI Conference on Human Factors in Computing Systems (CHI '26), April 13--17, 2026, Barcelona, Spain}
\acmDOI{10.1145/3772318.3791325}
\acmISBN{979-8-4007-2278-3/2026/04}

\begin{document}

\title{Memory Printer: Exploring Everyday Reminiscing by Combining Slow Design with Generative AI-based Image Creation}


\author{Zhou Fang}
\affiliation{%
\department{Industrial Design}
  \institution{Eindhoven University of Technology}
  \city{Eindhoven}
  \country{the Netherlands}}
\email{z.fang1@student.tue.nl}
\orcid{0009-0000-8305-997X}

\author{Janet Yi-Ching Huang}
\affiliation{%
\department{Industrial Design}
  \institution{Eindhoven University of Technology}
  \city{Eindhoven}
  \country{the Netherlands}}
\email{y.c.huang@tue.nl}
\orcid{0000-0002-8204-4327}

\renewcommand{\shortauthors}{Fang and Huang}

\begin{abstract}

Generative Artificial Intelligence (GAI) offers new opportunities for reconstructing these unrecorded memory scenes, yet existing web-based tools undermine users' sense of agency through disengaging and unpredictable interactions. In this work, we advance three design arguments about how slow, tangible interaction can reshape human–AI relationships by making temporality, embodied agency, and generative processes experientially legible. We instantiate these arguments by presenting \textit{Memory Printer}, a tangible design that combines silk-screen printing metaphors with text-to-image generation. The design features layered reconstruction that decomposes image generation into incremental steps, a physical wooden scraper enabling embodied control over image revelation, and built-in printing that produces tangible photos. We examine these arguments through a comparative study with 24 participants, exploring how participants engage with, interpret, and respond to this interaction stance. The study surfaces both opportunities—such as vivid memory evocation, heightened sense of control, and creative exploration—and critical tensions, including risks of false memory formation, algorithmic bias, and data privacy. Together, these findings articulate important boundaries for deploying generative AI in emotionally sensitive contexts.

\end{abstract}

\ccsdesc[500]{Human-centered computing~Interaction design}
\ccsdesc[500]{Human-centered computing~Empirical studies in interaction design}
\ccsdesc[500]{Human-centered computing~Systems and tools for interaction design}
\keywords{Generative Artificial Intelligence,
Tangible Interfaces,  Slow Design, Reminiscing Support}

\begin{teaserfigure}
  \includegraphics[width=\textwidth]{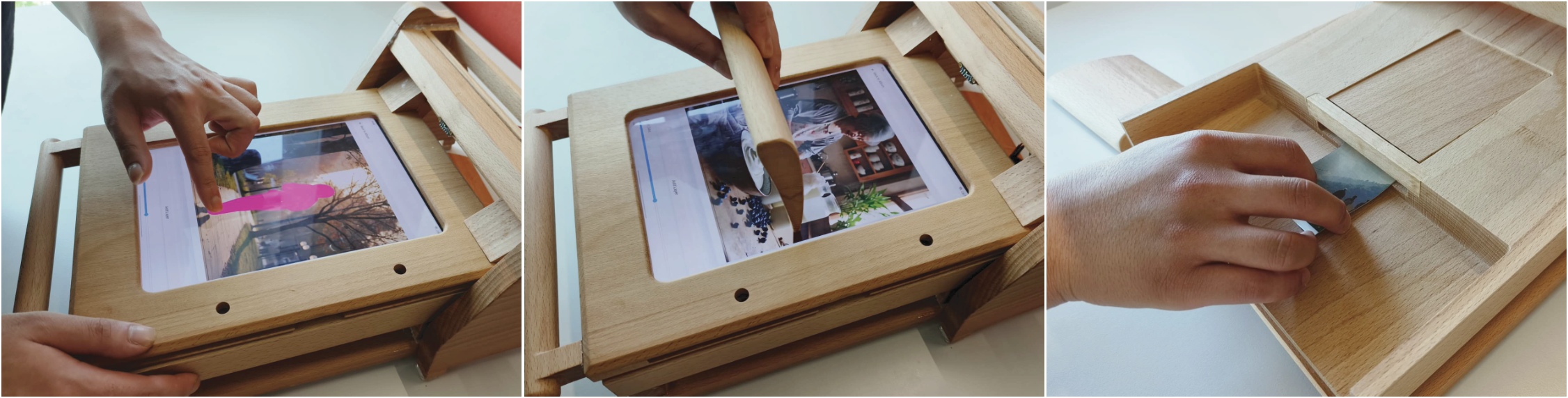}
  \caption{Memory printer that allows users to recreate their memory scenes using GAI and print them as memoryphoto.
}
  \label{fig:teaser}
\end{teaserfigure}
\maketitle

\section{Introduction}
Memory is one of the core components of the human cognitive system. It not only carries an individual's life experience and knowledge base, but also plays a key role in shaping self-awareness, emotional expression and decision-making \cite{Prebble2013}. The triggering of memory is very complex and involves a variety of internal and external factors. Research shows that the triggering of memory usually depends on the stimulation of memory cues. These memory cues can be physical factors in the external environment, or internal emotions and associations \cite{DanCosley2012}. Artifacts have long been used as an important carrier of memory. People use photos, diaries, souvenirs and other forms of artifacts to record memories. These memory triggers help individuals to reproduce past moments and even certain emotions through sensory impressions \cite{Tsai2018}. Based on this, in the field of health, people use these artifacts as therapeutic tools to alleviate certain cognitive diseases such as Alzheimer's disease \cite{Wilson1992, Cassel2015}. In everyday life, people share photos with their friends and family and tell each other stories from the past. This not only leads to reminiscing, but also promotes communication and strengthens emotional bonds \cite{Hoven2014}. 

Over the years, researchers have explored how to use technology-mediated artifacts to support reminiscing. With the development of Artificial Intelligence (AI), large language models (LLMs) have shown potential in human-like dialogue and information retrieval. This has led researchers to work on how to use AI to enhance reminiscing. At present, most research in this field is still based on LLM such as ChatGPT. There has been less research on how other types of AI models can enhance reminiscing. Furthermore, among various memory cues, visual stimuli, especially images, have always been considered one of the best memory cues for evoking memories because they can help individuals to visualize abstract events \cite{Eardley2006}. At the same time, the information is more direct and richer than that of text \cite{Zinko2019}. However, not all the important moments can be recorded in photos or other forms of media. Memories that are not recorded often fade, distort, and disappear over time, which makes it impossible for us to relive certain moments. How to effectively evoke these unrecorded memories has always been a challenge. Generative Artificial Intelligence (GAI), with its ability to quickly generate realistic images from textual descriptions, offers new opportunities to address this challenge. Yet the introduction of generative images into memory work raises questions not only about what AI can generate, but also about how people interact with and make sense of these generative processes.

To explore this potential, we conducted a preliminary study with eight participants to examine how users interact with existing web-based GAI tools to reconstruct personal memory scenes. The findings confirmed that GAI has potential to scaffold memory retrieval---participants recalled additional details through iterative image generation. However, this preliminary study also revealed a critical challenge: participants consistently reported a diminished sense of agency, describing the interaction as emotionally disengaging and the generation process as unpredictable and difficult to control.

These findings point to a fundamental tension: while GAI shows promise for memory reconstruction, existing web-based tools undermine the sense of personal control, agency and authorship that meaningful memory work requires. The sense of agency---the subjective experience of controlling one's actions and their outcomes \cite{limerick2014experience}---is supported by multiple cues, including motor signals from physical action and external feedback that enables reflection \cite{moore2012sense}. The screen-confined, automated nature of web-based GAI tools disrupt both sources: users cannot execute meaningful embodied actions, and rapid, unpredictable generation forecloses opportunities for reflection. We propose that two complementary design strategies can address these challenges. Slow design, which advocates for technology that creates space for reflection rather than prioritizing efficiency, aligns with both the reconstructive nature of memory and the need to restore user agency---memories emerge gradually through reconstruction \cite{Schacter2007}, and deliberate slowness creates temporal space for evaluation and sense-making. Tangible interaction can restore the embodied engagement that screen-based interfaces eliminate, providing richer agency cues through movement, proprioception, and spatial reasoning \cite{ishii1997tangible, dourish2001action}.

Based on these considerations, we follow the Research through Design methodology~\cite{zimmerman2007research} and Gaver's characterization of design exemplars~\cite{gaver2012should}. We advance three interconnected design arguments about how slow and tangible interaction might reshape human–AI interaction in the context of everyday reminiscing. Drawing inspiration from traditional screen printing---an inherently slow, embodied, and layered draft practice---we argue that (1) deliberate slowness creates space for reflection; (2) embodied interaction restores agency; and (3) decomposed control enables negotiation with AI. These arguments address a specific design space: emotionally significant content, autobiographical material where authorship matters, and contexts tolerating deliberate pace. In efficiency-critical domains, these arguments may not apply.

To stage these arguments for experiential exploration, we designed Memory Printer, a tangible design that combines traditional screen printing with generative artificial intelligence to support reminiscing in everyday scenarios. Memory Printer is not a solution to the challenges identified above; rather, it operates as a design exemplar---a materialized argument that instantiates these contestable propositions and invites their critique. The prototype enables users to gradually recreate memory scenes through a layered reconstruction process, decomposing prompt engineering into incremental steps. Users interact through a physical wooden `scraper' that controls the revelation of the generated images, replacing passive button-clicking with embodied action and allowing users to control the pace of image revelation. The prototype also prints the reconstructed memory as a tangible photograph, transforming digital interaction into a lasting physical artifact.

We conducted a user study with 24 participants comparing the Memory Printer with a web-based GAI tool. Our goal is to understand users' experience of using the Memory Printer and examine these design arguments empirically. Specifically, we examine how this interaction stance reshapes participants' sense of control, emotional resonance, and creative exploration during memory reconstruction. Our findings lend support to the proposition that Memory Printer's design qualities---slowness, tangibility, and layered control---facilitated more coherent memory evocation while enhancing participants' sense of control, emotional resonance, and creative exploration. While our findings demonstrate the potential of combining slow design with GAI for reminiscing, we also acknowledge that critical concerns---false memories, algorithmic bias, and data security---represent important boundaries and considerations for deploying GAI in emotionally sensitive contexts like reminiscing. 

This work contributes to HCI in two ways. First, we offer Memory Printer as a design exemplar that demonstrates how slow design principles and tangible interaction can work synergistically in human-AI interaction, providing a concrete--yet contestable---reference point for future design work. Second, through the process of designing, building, and evaluating Memory Printer, we generate experiential knowledge about the possibilities and pitfalls of GAI-supported reminiscence, contributing grounded insights rather than universal guidelines to an emerging design space.

\section{Related Work}
\subsection{Technology-Mediated Reminiscing in Everyday Contexts }
From photographs to diaries, letters to mementos, artifacts play a crucial role as external triggers for memory, evoking associations through sight, sound, smell, taste, and touch~\cite{Herz2004, Wei2019, Kim2022, Niemantsverdriet2016}.  

The effectiveness of artifacts in memory evocation has been studied and applied in various fields. In the health field, reminiscence therapy based on autobiographical memory is used to treat Alzheimer's disease. Stimulating memory through photos and personal items can help patients improve cognitive function \cite{Subramaniam2013}. Technological advances have expanded these possibilities for everyday use across diverse user groups  \cite{Lazar2014}. Recent HCI research has explored various approaches: MemoryReel \cite{Wei2019} encourages reflection through photos and sounds. Kim et al. designed an interactive photo frame as a memory aid \cite{Kim2022}. Memento \cite{Niemantsverdriet2016} is an interactive pendant, captures daily sound clips to stimulate memories unobtrusively. 

\subsection{Use AI-Powered Design Enhance Reminiscing}
Research using AI to directly support reminiscing is still in its infancy~\cite{Jeung2024}. Related research has pointed out the unique advantages of AI in some fields. For example, the active chat-bot memory reviver based on a visual recognition model and LLM has been used to help visually impaired people recall a photo collection \cite{Xu2024}. However, most approaches rely on LLM (e.g., ChatGPT), which emphasizes the importance of dialog with users. With the iterative development of generative adversarial networks (GANs), it has become possible to quickly generate images and even videos, some of which are almost indistinguishable from the real \cite{Karras2019}. This provides more possibilities for using memory cues. Previous research has shown that the combination of visual and even tactile multi-sensory experiences is conducive to enhancing the emotional impact of memories, thus better evoking memories \cite{Yamauchi2018}. 
\subsection{Slow Design and Reminiscing}
Memory is a cognitive process that is susceptible to emotional and cognitive factors~\cite{Conway2000}. It is not simply a static storage system, but a dynamic process that is reconstructive. Memory can be affected by many factors, such as personal cognition, memory behavior, and the environment~\cite{Schacter2007}. This means that memories often undergo a long and slow process of remodeling after they are formed. The theory of slow design suggests that technology can be made slow in various ways because it takes time for reflection~\cite{Hallnaes2001}, which presents a symbiotic relationship with the characteristics of memory. Slow design provides an environment in which time is stretched out, allowing users to think. This gradual, reflective engagement allows users to develop a deeper connection with their memories over time, rather than obtaining fleeting, transient moments of recollection. HCI research has explored this relationship through various prototypes. Photobox~\cite{Odom2014} randomly prints users‘ photo collections every month. This slow, unpredictable interaction  promotes more memories and reflections during the waiting. Photoclock~\cite{Chen2023} displays photos taken at the same time in the past. This design helps users reflect on the past and establish an evolving relationship between the user and their photo collection. Similar explorations include Olly~\cite{Odom2019} and memory tracer~\cite{White2023}, all these projects explore how slow design supports ongoing, reflective memory engagement. Our work attempts to further explore the positive impact of slow design on reminiscing based on new technologies. 

\subsection{User Agency in Human-AI Interaction}
\label{subsec:User Agency in Human-AI Interaction}
\subsubsection{Theoretical Background}
Human agency, the subjective experience of controlling one's actions and their consequences, has long been a central concept in the field of HCI. Limerick et al. \cite{limerick2014experience} argue that as an increasing number of agentic interactions in our daily lives involve technology, understanding how interface design affects users' sense of agency is crucial for human-computer interaction research. 
Previous studies have established several models to explain how humans develop a sense of agency. The comparator model suggests that agency arises when the predicted outcomes of our actions match the actual outcomes we perceive \cite{frith20162000}. Building on this, Moore and Fletcher \cite{moore2012sense} proposed the cue integration\textit{ approach}, arguing that the sense of agency is not determined by any single factor but emerges from multiple cues.
The HCI community has developed several design paradigms that effectively support user agency. Shneiderman's work \cite{shneiderman1983direct} established that physical actions and incremental operations can enhance users' sense of control. Dourish's work \cite{dourish2001action} further emphasized how bodily engagement with computational systems creates richer experiences of agency than purely symbolic interaction. Tangible interfaces extend this principle by coupling digital information to physical objects, providing users with haptic feedback and proprioceptive awareness that screen-based interfaces cannot offer \cite{ishii1997tangible}.

\subsubsection{Agency Disruption in Human-AI Interaction}
Prior work noted that as more task execution stages involve automated processing, users' sense of control diminishes even when system performance improves \cite{berberian2012automation}. This paradox becomes more pronounced when AI systems make autonomous decisions or generate content without transparent, incremental user control. When users interact with these GAI tools, their actions are reduced to inputting prompts and clicking buttons—operations that provide minimal sensorimotor feedback. While this intent-based outcome specification represents a powerful new interaction paradigm \cite{nielsen2023ai}, it fundamentally transfers control from users to opaque generative models. 

Researchers have begun to explore how generative AI specifically disrupts user agency. Recent research \cite{mahdavi2024ai} found that users struggle to cope with generative models' unpredictability—minor variations in prompts can yield drastically different outputs. This lack of transparency and predictability undermines two key prerequisites for agency established in early HCI research: understanding system behavior and being able to reliably produce intended results \cite{limerick2014experience}. Moreover, the cognitive burden imposed by "prompt engineering" shifts users' attention from creative goals to technical problem-solving, potentially diminishing their sense of ownership over generated content.
Given these challenges, researchers have explored various strategies to preserve or restore user agency in AI-assisted environments. Some work emphasizes transparency and explainability to achieve this goal \cite{miller2019explanation}, while others attempt to distribute control across multiple stages of the AI process \cite{louie2020novice}. Despite some progress, most existing solutions remain screen-based and focus on single dimensions of agency (e.g., execution control through iterative prompt refinement). Prior research suggests that through multisensory integration, physical manipulation can provide richer agency cues \cite{coyle2012did}. However, generative AI tools largely abandon physicality in favor of purely digital, screen-based interfaces optimized for efficiency rather than experiential richness. There remains limited research on how factors such as temporality, materiality, and embodied cognition work synergistically to support multidimensional agency in generative AI systems.
Our research also aims to explore how tangible and slow interaction design can support agency in human-AI interaction. Through user testing with Memory Printer, we provide empirical evidence that these design strategies can enhance users' sense of agency beyond what purely screen-based tools can achieve, even when the underlying generative process remains opaque.

\section{Preliminary Study}
To inform the design of the Memory Printer, we conducted a preliminary study with two primary objectives: (1) to validate whether generative GAI can effectively assist users in reconstructing memory scenes, and (2) to identify specific limitations in existing web-based GAI tools that prevent users from better reconstructing memory scenes. This formative study provided empirical grounding for our design principles and revealed critical gaps in how conventional AI interfaces support reflective memory practices.
\subsection{Participants and Procedure}
We recruited eight participants (four women and four men) through social media advertising. All participants were between 20 and 30 years old. Our goal in this preliminary study was to observe how users interact with web-based GAI tools for memory reconstruction and to identify challenges in this process, rather than to specifically study novice users.

\begin{figure*}[t]
    \centering
    \includegraphics[width=1\textwidth]{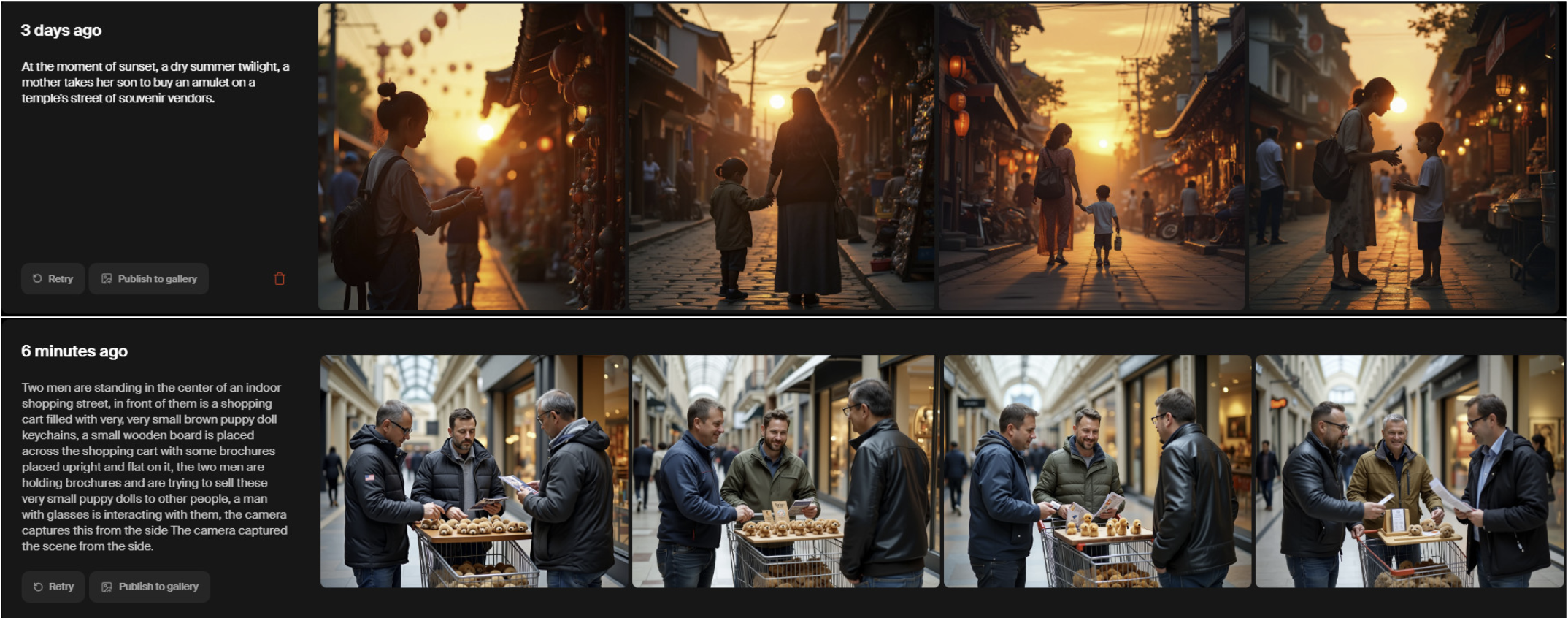}
    \caption{Participants attempted to use KreaAI to recreate scenes from their memories.
}
\label{fig:krea}
    \Description{2}
\end{figure*}

The testing consisted of three phases: First, participants described their personally experienced events . Next, they used AI to generate images that could reconstruct these events. For this purpose, we employed KreaAI, a web-based GAI tool. During this phase, participants iteratively refined their prompts based on their descriptions until they believed the generated images matched the scenes in their memory (Figure~\ref{fig:krea}). Finally, we conducted semi-structured interviews. We aimed to collect qualitative data from two perspectives: (1) whether participants believed the AI accurately recreated their scene descriptions, and (2) participants' opinions about this web-based AI image generation tool.

\subsection{Findings}
\subsubsection{GAI Tool Shows Potential for Assisting in Memory Scene Reconstruction}
After multiple prompt adjustments, all participants successfully generated images they believed could represent their memories. During the interaction, participants formed a behavioral loop of ``recalling events $\rightarrow$ modifying prompts $\rightarrow$ observing generated image details $\rightarrow$ recalling more scene details.'' When observing generated images, participants proactively compared the images with their memories and modified the image details. This behavior not only gradually aligned the images with their memory scenes, but Participants also felt their memories gradually becoming clearer, with details growing richer. However, this process was cognitively demanding. Participants needed to recall memory details, translate them into effective textual prompts, evaluate generated images for accuracy, and iterate based on discrepancies. This imposed an additional cognitive burden on participants.
\subsubsection{Interaction with Web-based GAI Tool Lacks Engagement}
Participants universally found prompt input difficult—despite careful selection, generated images often mismatched expectations, leading to frustration and lost interest. Participants' frustrations extended beyond mere difficulty. They perceived the entire interaction as ``emotionally flat and disengaging.'' The process consisted only of repetitive typing and mouse clicking, lacking appeal. 

\subsubsection{Participants Lacked Sense of Control over Image Generating}
Image generation uncertainty further hindered participants' ability to obtain the desired images. Participants found that when attempting to modify images, they struggled to control specific details precisely. Moreover, when they adjusted their prompts to fix certain elements, other aspects they were previously satisfied with often changed unexpectedly. This inscrutable, fully automated process left them feeling frustrated. Although participants were generally satisfied with the overall atmosphere of the generated images, this unpredictability made them feel a lack of control. They agreed that it prevented them from building trust in AI and hindered deeper reflection on their memories.

\subsection{Design Principles}
Based on the findings, we derived three design principles to guide prototype's development. Each principle addresses a specific dimension that was compromised in existing web-based GAI tools. Through these principles, we aim to create more engaging and meaningful interactions that better support users' reminiscing and reflection.

\subsubsection{Create Interaction Experiences beyond Screen}
Our findings indicate that participants described this screen-based, simple yet repetitive interactive process as a bland and disengaging experience. Meanwhile, memory is personal, emotionally charged, and meaningful. To better support reminiscing, interactions should engage users beyond mere efficiency. Previous research explored how physical objects can transform users' interaction experiences. To better understand user needs, we applied Norman's ``visceral, behavioral, and reflective'' framework during the prototype design process~\cite{Norman2007}. We aim for a GAI tool that effectively supports reminiscence while evoking feelings associated with memory---cultivating a retro, warm atmosphere that strengthens users' emotional connection to the events they want to reconstruct.

\subsubsection{Provide Guidance for Users' Memory Recall}
Research indicates that memory details tend to emerge in a particular order: either by association from salient details or by temporal sequence~\cite{Conway2000}. In the preliminary study, generated images supported this process by serving as visual cues for recalling memory details, but did not guide the sequential emergence of these details—users had to address all aspects of memory reconstruction at once. At the same time, while GAI can support this process, it imposes substantial cognitive demands. We attempted to decompose the entire  process into few phases. This phased approach serves a dual purpose: (1) reducing cognitive burden by breaking complex tasks into manageable steps; (2) To provide natural reflection points through incremental, segmented interactions, allowing users to pause, evaluate progress, and progressively elicit deeper memories.

\subsubsection{Enhance Users' Sense of Control over Generated Images}
The purely virtual nature of web-based GAI tools fundamentally undermines users' sense of control. Moreover, when entire images transform based on prompt modifications, users cannot discern which specific prompt elements caused particular visual changes---their actions fail to produce clearly attributable effects. This resulting unpredictability leaves users feeling constrained by an incomprehensible system. Previous research indicates that sense of agency is influenced by multiple cues, including internal motor signals and external contextual cues~\cite{moore2012sense}. By providing users with clear, attributable methods of control, we aim to restore their sense of agency over the generative process through these two aspects.

\section{Memory Printer}
Based on the insights from our preliminary study, we designed Memory Printer, a tangible interface inspired by traditional screen printing combined with generative AI capabilities to support everyday reminiscing.  The prototype consists of a wooden housing, a physical scraper for controlling image revelation, a layered redrawing interface, and a built-in Zink printer that produces tangible photo outputs (Figure~\ref{fig:teaser}). 

Memory Printer is not merely a functional tool---it embodies a design-as-argument approach. Following Research through Design methodology~\cite{zimmerman2007research} and Gaver's characterization of design exemplars~\cite{gaver2012should}, we see the Memory Printer as a materialized argument: a designed object that embodies contestable theoretical propositions about human–AI interaction and stages them for experiential evaluation in everyday use. We do not claim that Memory Printer is definitive; instead, it operates as a design exemplar that instantiates and invites critique of these claims. Specifically, we advance three design arguments, each operationalized in specific design features and examined in our empirical study.

\subsection{Design Inspiration: Screen Printing as Metaphor}

The design of the Memory Printer draws inspiration from screen printing, an ancient technique where a scraper pushes pigment through stenciled templates, building complete images through successive color layers. This craft practice suggested three design possibilities that address the limitations we observed in web-based GAI tools.

First, screen printing is inherently slow and deliberate. Each layer requires preparation, alignment, and careful execution—rushing produces poor results. This contrasts sharply with the instantaneous generation of web-based tools, suggesting that enforced slowness might create a valuable reflective space.

Second, screen printing is fundamentally embodied. Practitioners grip the scraper, feel its resistance against the screen, and control pressure and speed through bodily skill. This physical engagement stands in stark contrast to the disembodied clicking of web interfaces.

Third, screen printing decomposes images into independent layers. Rather than producing complete images in one action, practitioners build complexity incrementally, addressing one color or region at a time. This layered construction enables systematic refinement impossible when outputs appear wholesale.

Beyond these functional parallels, screen printing's vintage aesthetic evokes nostalgia appropriate for memory work, and its familiar physicality reinforces the embodied engagement central to our second design argument. The conceptual alignment runs deeper still: spreading activation theory~\cite{Foster2017} suggests that memories emerge sequentially with salient details first, while self-memory system theory~\cite{Conway2000} describes hierarchical retrieval where details integrate progressively---processes that mirror screen printing's layered image construction. 

\subsection{Design Arguments}
Drawing on these inspirations, we advance three interconnected arguments about how tangible, slow interaction can reshape human-AI relationships in emotionally sensitive contexts. Each argument transforms an observed quality of screen printing into a testable proposition about human-AI interaction:

\subsubsection{Argument 1: Deliberate Slowness Creates Space for Reflection}
Contemporary GAI tools prioritize immediacy, assuming that faster is better. We argue that for emotionally significant content like personal memories, enforced slowness enables deeper engagement that rapid interaction forecloses. When users must invest time to see results, they gain opportunities to prepare emotionally, evaluate incrementally, and reflect on emerging content. Memory Printer operationalizes this through the scraper mechanism, which requires deliberate physical movement to reveal images, and through the layered modification process, which decomposes rapid iteration into sequential stages of observation, drawing, description, and evaluation.

\subsubsection{Argument 2: Embodied Interaction Restores Agency}
Web-based GAI tools confine users to typing and clicking, contributing to a diminished sense of control. We argue that physical manipulation---gripping, pushing, feeling resistance---provides proprioceptive and haptic feedback that creates richer agency cues. Memory Printer operationalizes this through the wooden scraper requiring grip and force, the warm birch wood housing, and the physical photo output that transforms digital interaction into a lasting material form.

\subsubsection{Argument 3: Decomposed Control Enables Negotiation}
Most GAI systems present complete outputs that users accept or reject wholesale, preventing users from understanding cause-effect relationships. We argue that decomposing generation into addressable components enables iterative negotiation with AI. Memory Printer operationalizes this through layered redrawing, where users create independent layers targeting specific image regions, each with its own mask and description, allowing attribution of specific changes to specific actions.

These arguments address a specific design space: content carrying emotional significance, autobiographical material where authorship matters, and contexts tolerating deliberate pace. In efficiency-critical domains, these arguments may not apply. By articulating them explicitly, we invite critical evaluation through our empirical study.

\subsection{Prototype Features}
Figure~\ref{fig:interaction} illustrates the complete interaction flow. Users first narrate initial memory impressions; the built-in microphone captures these and generates a preliminary base image. Users then examine the image and modify details through layered redrawing. Finally, they print the reconstructed memory as a physical photograph.
\begin{figure*}[t]
    \centering
    \includegraphics[width=1\textwidth]{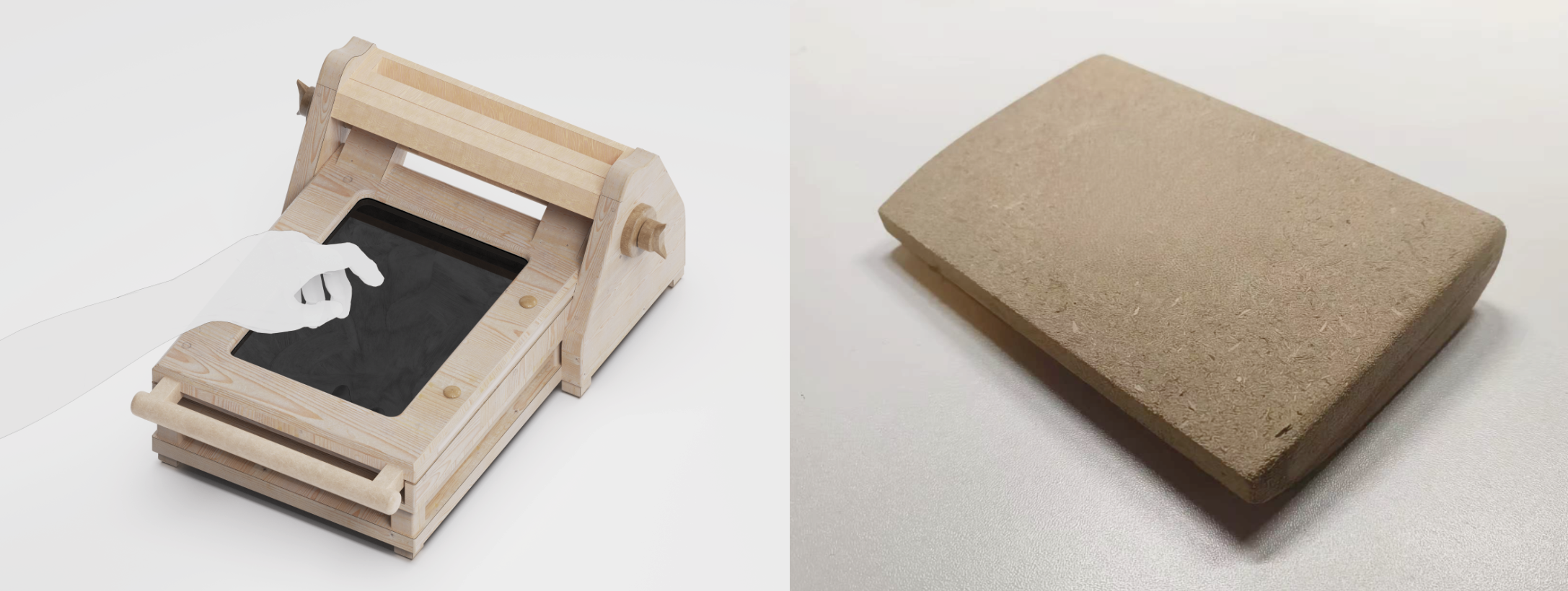}
    \caption{The memory printer and the wooden scraper
    \label{fig:printer-scraper}
}
    \Description{2}
\end{figure*}

\begin{figure*}[t]
    \centering
    \includegraphics[width=1\textwidth]{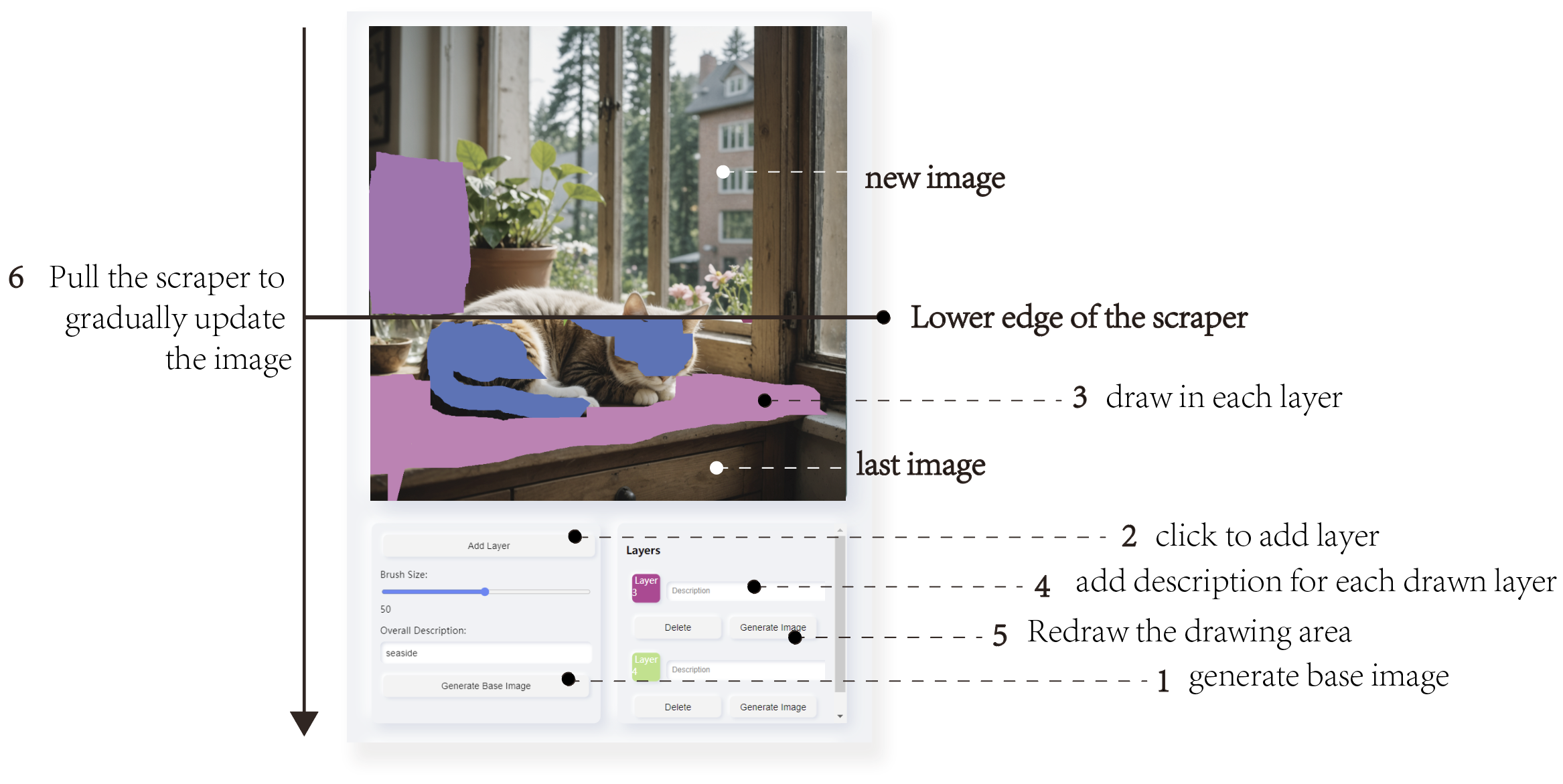}
    \caption{Six steps to restore memory scenes using the memory printer}
    \Description{2}
    \label{fig:interaction}
\end{figure*}

\begin{figure*}[t]
\centering

\begin{minipage}{0.48\textwidth}
    \centering
    \includegraphics[width=\linewidth]{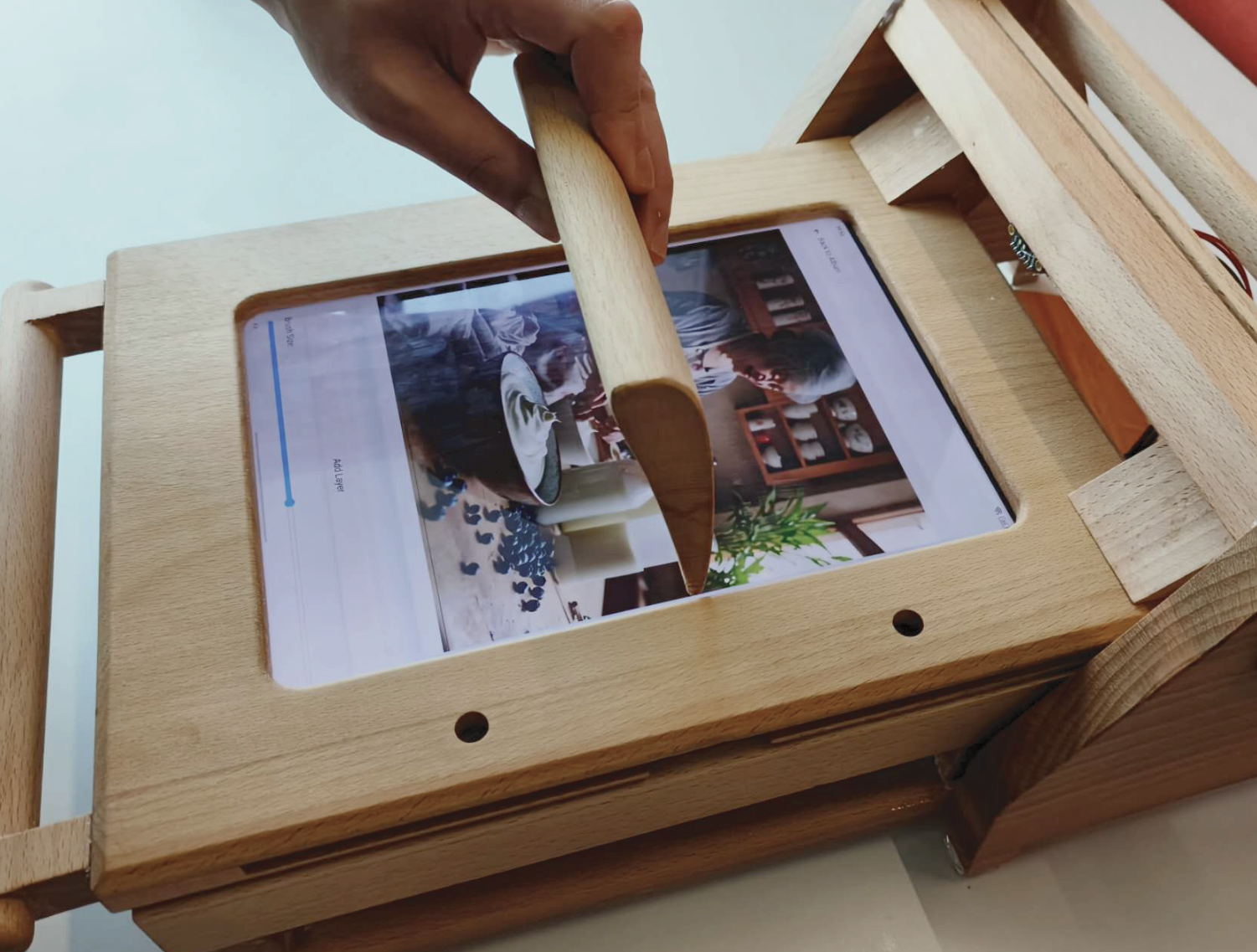}
    \caption{The image is gradually updated as the scraper moves.}
    \label{fig:scraper}
\end{minipage}
\hfill
\begin{minipage}{0.48\textwidth}
    \centering
    \includegraphics[width=\linewidth]{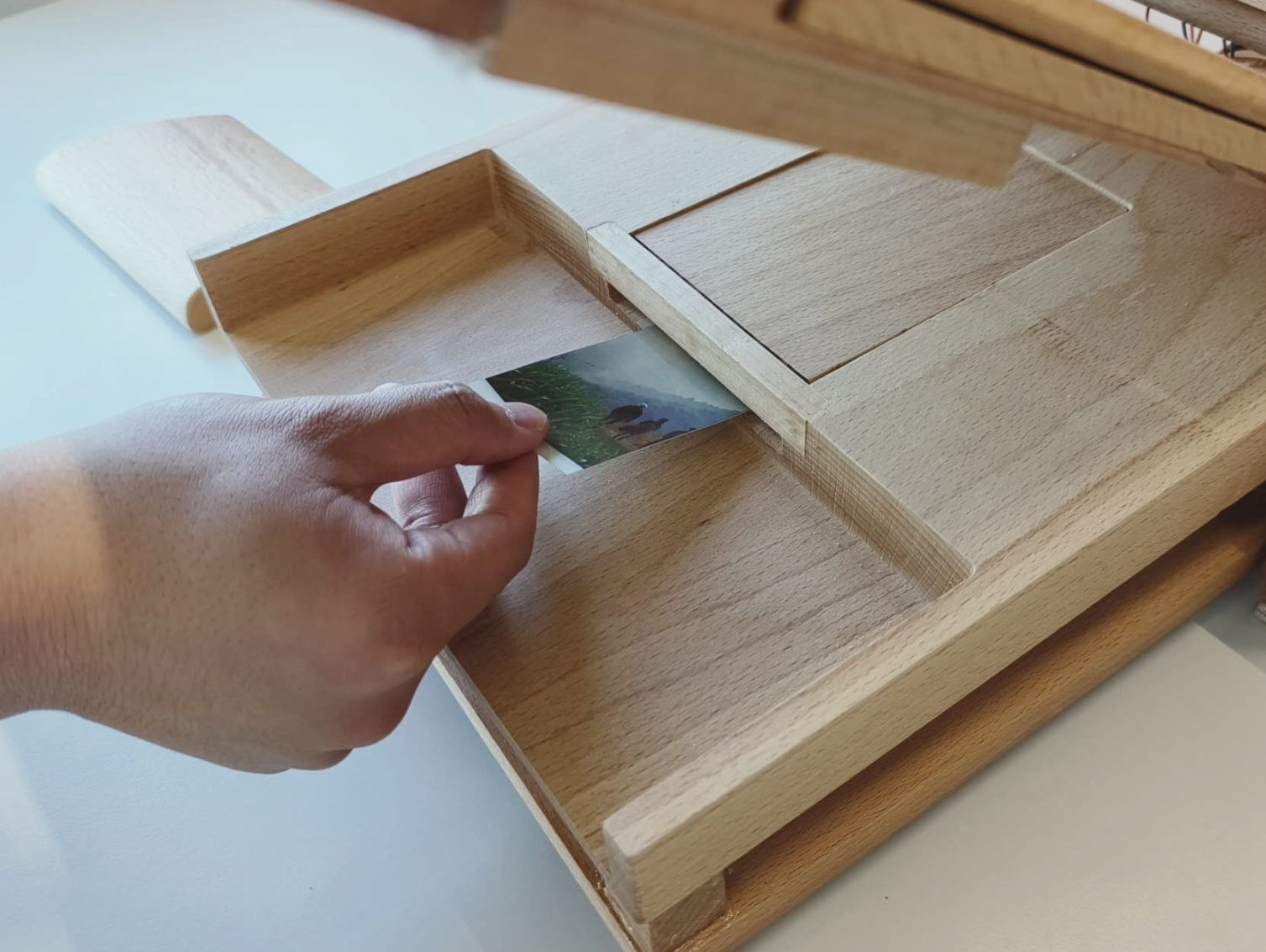}
    \caption{User retrieves images printed by Zink printers from the memory printer.}
    \label{fig:printing-photo}
\end{minipage}

\end{figure*}

\subsubsection{Scraper-Controlled Image Revelation}
During image generation, we introduced a ``scraper'' as the interactive tool for image control. We employ a distance sensor to monitor the scraper's movement. As the scraper moves along the screen toward its bottom edge, the modified image gradually reveals itself in tandem with the scraper's motion, with the image's bottom edge always aligned with the scraper (Figure~\ref{fig:scraper}). This approach transforms the web-based GAI tool, which relies solely on button clicks or other virtual interactions for image generation. The scraper not only makes the entire process more akin to actual screen printing but also allows users to control image presentation at their own pace, further enhancing their sense of control. As users repeatedly move the scraper, they can carefully compare pre- and post-modification versions of the image. This comparative method enables users to more intuitively evaluate whether local changes in the image align with their expectations.


\subsubsection{Layered Redrawing}
\begin{figure*}[t]
    \centering
    \includegraphics[width=1\textwidth]{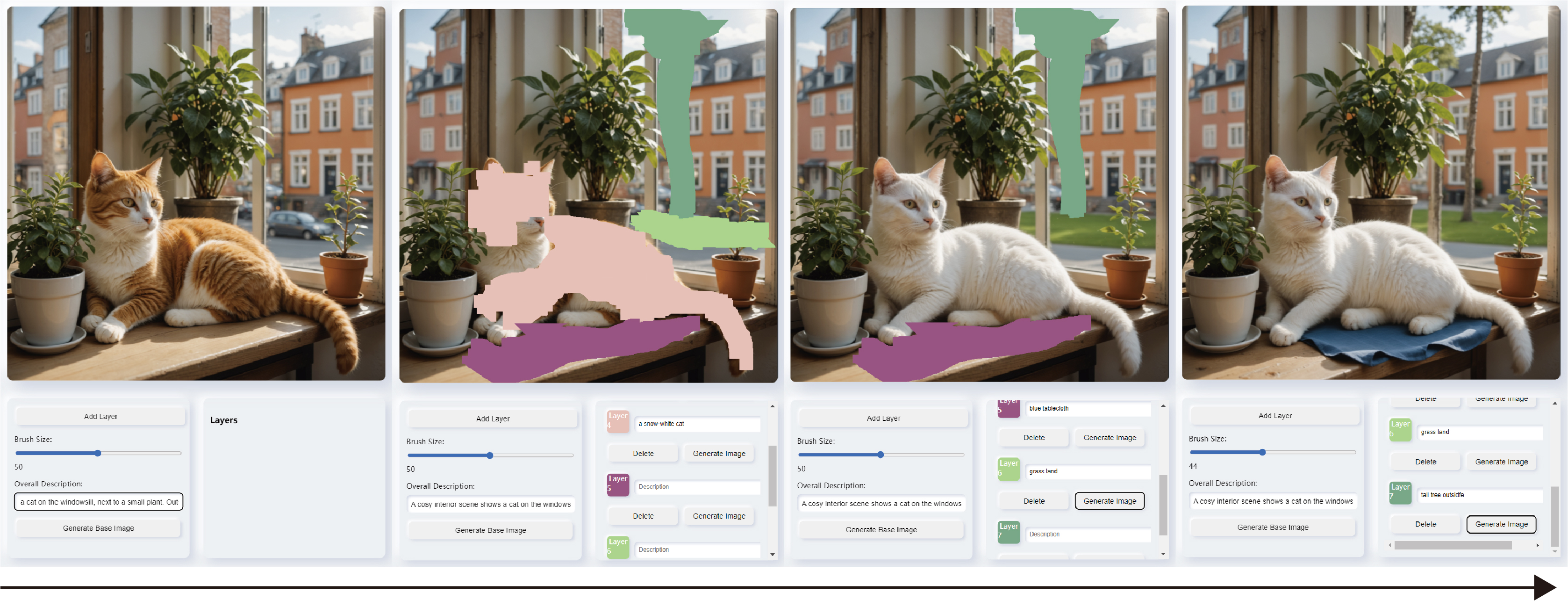}
    \caption{Demonstration of the process of modifying and redrawing parts of an image. Users first create a base image generated by AI. Then add layers to the image, drawing within each layer and adding prompts for the drawn parts. Subsequently, based on the drawn content and prompts, the base image is redrawn layer by layer.
}
    \Description{2}
    \label{fig:layered-redrawing}
\end{figure*}

Rather than regenerating entire images from revised prompts, users modify specific regions through independent layers. For each layer, users draw directly on the screen to create a mask indicating the target region, then provide verbal descriptions of the desired changes. The system redraws only the masked area based on these descriptions (Figure~\ref{fig:layered-redrawing}). This decomposition serves two purposes: reducing cognitive load by focusing attention on one element at a time, and enabling causal attribution—users can identify which layer produced which change. Users can create multiple layers addressing foreground, middle ground, and background elements separately, building systematic modification strategies.

\subsubsection{Physical Photo Output}
When satisfied with the reconstructed image, users lift the upper portion of the device to activate the built-in Zink printer, which produces a tangible photograph, ``a photo from the past'' (Figure~\ref{fig:printing-photo}).  By doing this, the digital Human-AI interaction process is transformed into a touchable photo, imbuing the memory reconstruction activity with deeper meaning. It serves not only as a copy of the final generated image but also as a witness and memento of the entire recollection journey. Users can preserve these printed photos as tangible mementos, filling gaps in personal histories lacking photographic documentation. Simultaneously, the deliberately slowed-down manual printing process creates a final space for reflection, allowing users to revisit the entire interaction, contemplate how they gradually reconstructed the memory, and imbue the memory printer's use with a sense of ritual.

\section{Empirical Study}

\subsection{Participants}
We recruited 24 participants through social media advertisements and snowball sampling. Before the user testing began, we ensured that all participants understood the testing objectives; confirmed that participants had no major illnesses, sensory impairments, and maintained clear thinking. This user study protocol was approved by the Institutional Review Board at the authors' institution, and all participants signed informed consent forms prior to participation. Participants were informed of the research purpose, procedures, potential risks (including exposure to potentially disturbing AI-generated content), and their right to withdraw from the study at any time.
The sample included 8 females (33\%) and 16 males (67\%), with ages ranging from 22 to 57 years (mean age M = 27.6 years). 23 participants were aged between 22 and 31 years, with one participant aged 57 (P17). Participants were primarily drawn from university communities and their relatives.
We did not require participants to have prior experience with generative AI tools. Among the 24 participants, 18 (75\%) reported having previously used AI image generation tools on various platforms (e.g., generating images using ChatGPT). Most of them described their experience as occasional or outcome-oriented rather than involving sustained, iterative interaction. The remaining 6 participants (25\%) reported no substantive prior hands-on experience with AI image generation. This variation in prior experience was incidental rather than a primary recruitment criterion; our central research objective was to examine the three design arguments (slowness, embodiment, and decomposed control) rather than to compare novice and experienced users. Participants were primarily drawn from university communities, resulting in a sample predominantly composed of young and highly educated individuals. We discuss the implications of this sample composition in the limitations subsection.


\subsection{Procedure}
\begin{figure*}[t]
    \centering
    \includegraphics[width=1\textwidth]{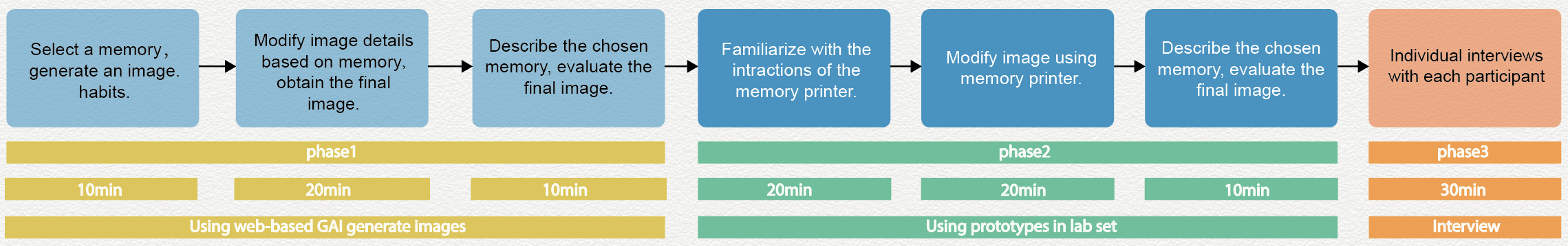}
    \caption{Procedure of the empirical study}
    \label{fig:test}
    \Description{2}
\end{figure*}

Our user study consisted of three stages, as shown in Figure~\ref{fig:test}. All user study was conducted in laboratory settings. In the first stage of the user study, participants were asked to select a memory and use a web-based GAI tool (KreaAI) to generate an image that could represent it. Within 20 minutes, they modified image details based on their recollections to make the images as close as possible to their memories. Afterward, participants shared this memory verbally while watching the image, and provided self-evaluations of the generated image. In the second stage, we introduced the memory printer to participants, demonstrated its workflow, and gave participants ten minutes to familiarize themselves with it. Participants were then asked to select another memory and attempt to use the memory printer to recreate memory scenes within 20 minutes. Similarly, after the generation activity concluded, participants introduced this memory and provided self-evaluations of the generated image. In both the first and second stages, all participants were required to think-aloud to help the research team better understand participants' behaviors and thought processes. In the third stage, we conducted one-on-one semi-structured interviews with participants. All participants were interviewed in English.

\subsection{Data Collection}
During the user tests, we recorded participants' testing procedures through audio and video recording . During the semi-structured interview phase, we used audio recording for documentation. Additionally, we preserved all AI-generated images produced during participants' testing processes. To supplement details missed by the above media, we used observational notes to specifically record participants' noteworthy behaviors and expressions. All collected audio recordings were transcribed using professional audio transcription software and saved in text format, with final proofreading conducted by the research team.

\subsection{Data Analysis}
We employed thematic analysis to synthesize and analyze the collected data, combining data-driven inductive and question-driven deductive approaches. This part of the work followed the step-by-step guide proposed by Braun \& Clarke \cite{Braun2006}. Specifically, our qualitative analysis included the following steps: (1) We carefully read all textual transcripts, re-watched all the videos, and supplemented missing details in the transcripts based on the recorded content. (2) We generated initial codes for each transcript through inductive coding. (3) We used affinity diagramming to organize codes, iteratively refining them to identify patterns and develop themes. (4) We reviewed and defined the final themes, extracting relevant quotes from the data and further categorizing them to understand how GAI helps participants construct memory scenes, and what utility our prototype provides in this process.

\section{Findings}
Our findings are organized around two overarching themes: how GAI-generated images support reminiscing activities (section~\ref{subsec:GAI-images-enhance-reminiscing}), and how Memory Printer's design arguments performed when evaluated through user experience (section~\ref{subsec:impact-tangible-interface}). We also report challenges and concerns that emerged (section~\ref{subsec:challenges-concerns}).

\subsection{GAI-Generated Images Enhance Reminiscing}
\label{subsec:GAI-images-enhance-reminiscing}
AI-generated images supported participants' recall activities in three ways: providing visual cues that made memories more concrete, helping participants organize fragmented recollections into coherent narratives, and evoking unexpected reflections through serendipitous image elements. While these characteristics appeared in both web-based GAI tools and Memory Printer, participants reported stronger effects with Memory Printer.

\subsubsection{AI-Generated Images as Visual Cues}
\begin{figure*}[t]
\centering

\begin{minipage}[t]{0.24\textwidth}
  \centering
  \includegraphics[width=\linewidth]{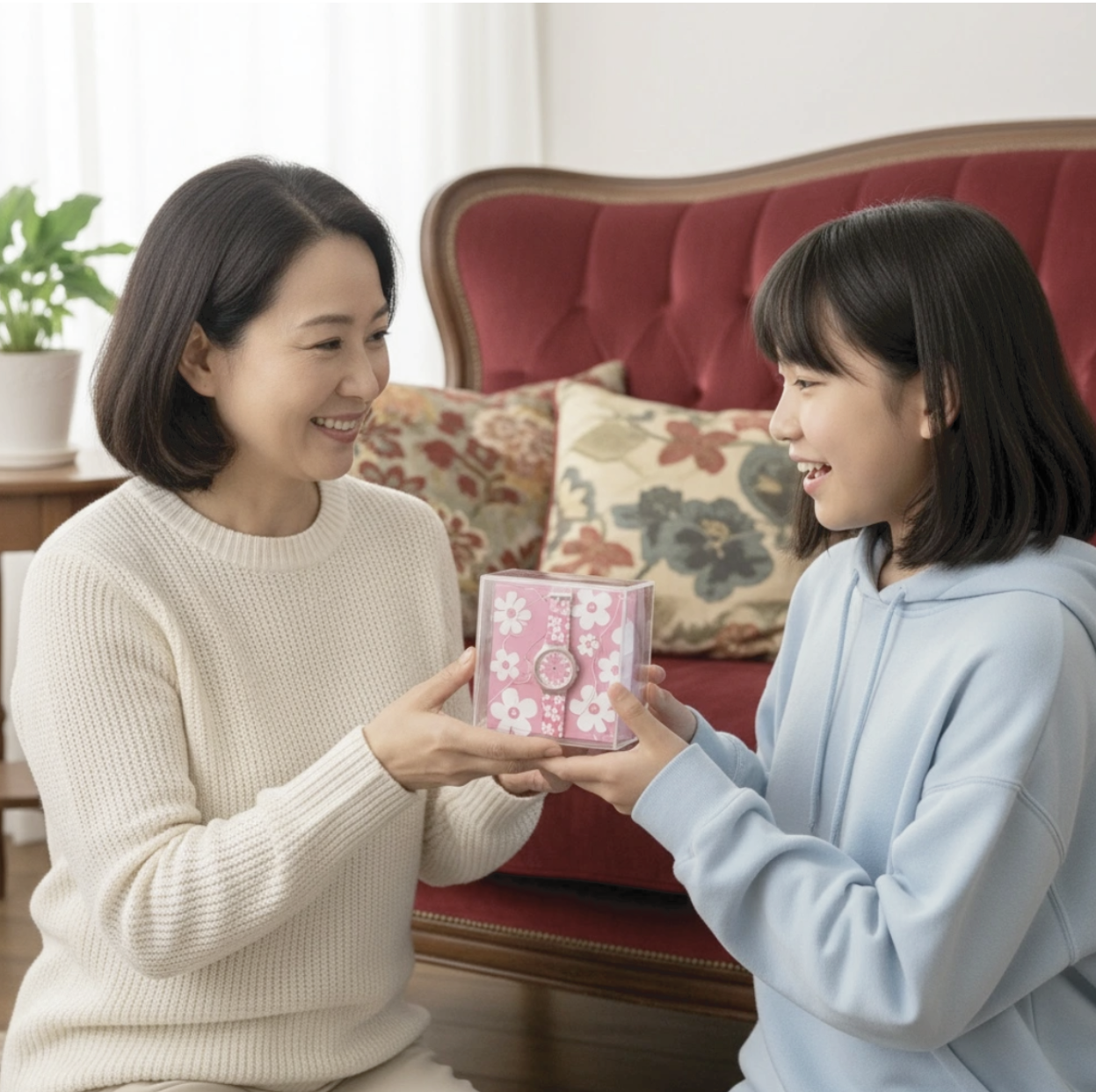}
  \captionof{figure}{The scene of P6's mother giving her a watch.}
  \label{fig:p6-watch}
\end{minipage}\hfill
\begin{minipage}[t]{0.24\textwidth}
  \centering
  \includegraphics[width=\linewidth]{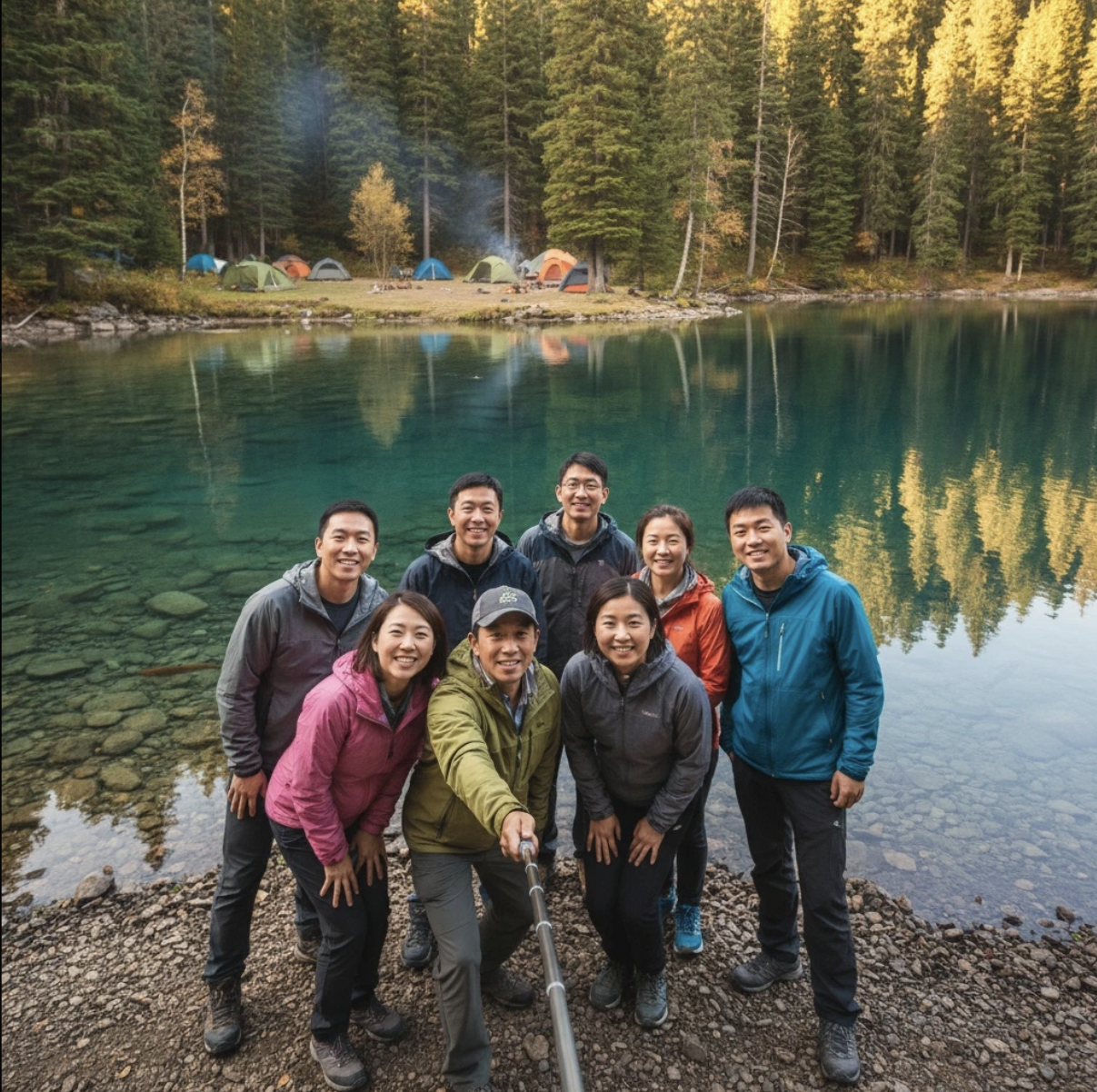}
  \captionof{figure}{Image from the first phase: P5 hiking with his friends.}
  \label{fig:p5-hiking}
\end{minipage}\hfill
\begin{minipage}[t]{0.24\textwidth}
  \centering
  \includegraphics[width=\linewidth]{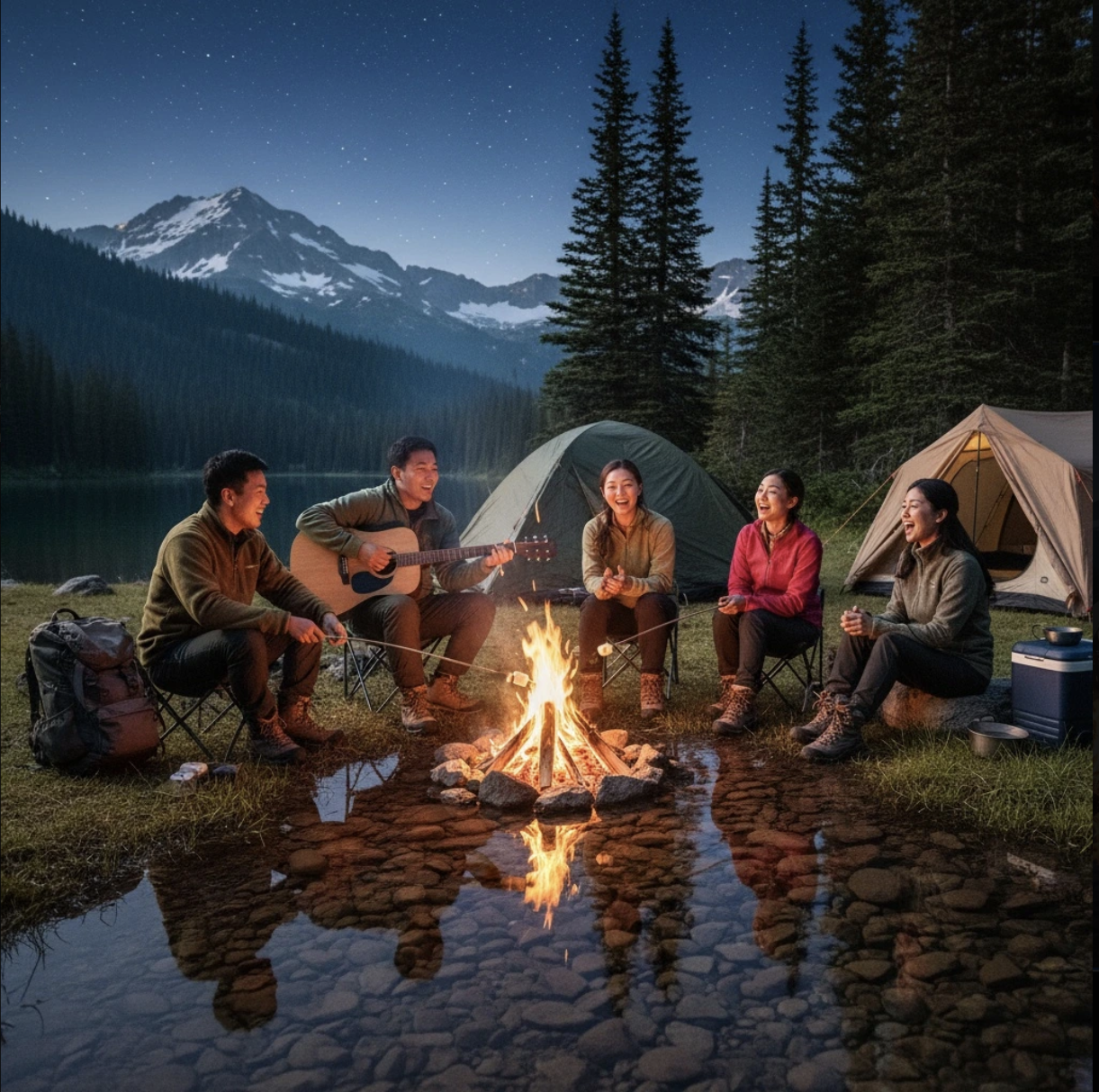}
  \captionof{figure}{Image from the second phase: P5 gathering around a bonfire with friends.}
  \label{fig:p5-bonfire}
\end{minipage}\hfill
\begin{minipage}[t]{0.24\textwidth}
  \centering
  \includegraphics[width=\linewidth]{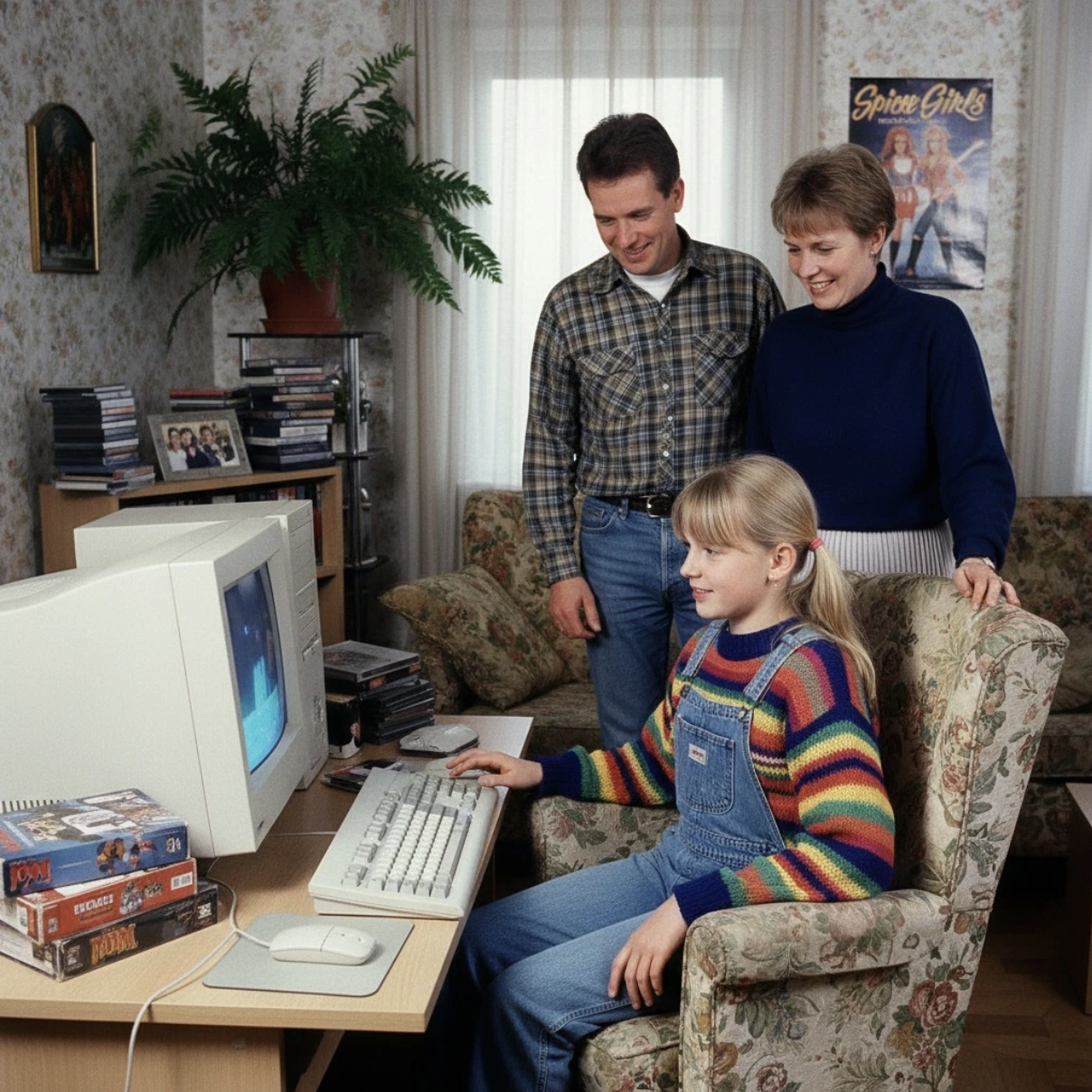}
  \captionof{figure}{P2's first experience playing video games as a child.}
  \label{fig:p2-games}
\end{minipage}

\end{figure*}

Participants used GAI tools to generate visual references for memories lacking photographic documentation. Throughout testing, participants progressively modified the generated images based on their recollections, which in turn provided increasingly relevant visual anchors. P6 attempted to recreate the scene of her mother giving her a watch during the second phase (Figure~\ref{fig:p6-watch}). In her interview, she stated: \textit{``I often think about this memory, but there are no photos of it...''} P6 viewed the AI-generated image as compensating for a memory scene lacking photographic recording, finding image-assisted recollection more vivid than unaided recall. 
Other participants concurred. P14 noted:\textit{``Although the AI-generated image doesn't perfectly match the memory, the atmosphere is very close.''} Participants naturally compared the details of the generated image with their own memories, repeatedly ``scanning'' their recollections. As they gradually aligned images with memories, their recollections deepened. P10 remarked during testing: \textit{``I remember this part wasn't like this—it should be... I rarely thought about these things before...''} While generating ideal images proved challenging at times, most participants (19 out of 24) sought not perfect visual matches but atmospheric resonance. P19 reflected: \textit{``The person in the image doesn't look like me, but they seem happy. I like this vibe.''}

\subsubsection{AI-Generated Images Help Organizing Memories}
Participants employing the Memory Printer (Phase 2) produced more detailed, narrative-driven accounts compared to their web-based tool experiences (Phase 1). P5 described travel experiences with friends in both phases. In Phase 1 (Figure~\ref{fig:p5-hiking}), P5 focused on the observed scenery, \textit{``the weather was nice''}, \textit{``the lake water was crystal clear.''} In Phase 2 (Figure~\ref{fig:p5-bonfire}), P5 provided richer descriptions with emotional commentary: \textit{``...the campfire made me so excited, and I was happy we were all together...''}---despite considering both memories equally pleasant and nostalgic. 
P22 finds this contrast striking. He noted that although both phases involved selecting memories of everyday moments with friends, Memory Printer evoked a deeper sense of immersion: \textit{``Interacting with the memory printer reminded me of the old days... I miss my friends... I didn't feel that way earlier (Phase 1).''} This suggests that Memory Printer helped participants construct more coherent memories, prompting reflection on additional details. AI-generated images helped participants narrate their stories, while enhanced storytelling led to more vivid memory details---a mutually reinforcing process.

\subsubsection{AI-Generated Images Evoke Reflection}
\begin{figure*}[t]
\centering

\begin{minipage}[t]{0.24\textwidth}
  \centering
  \includegraphics[width=\linewidth]{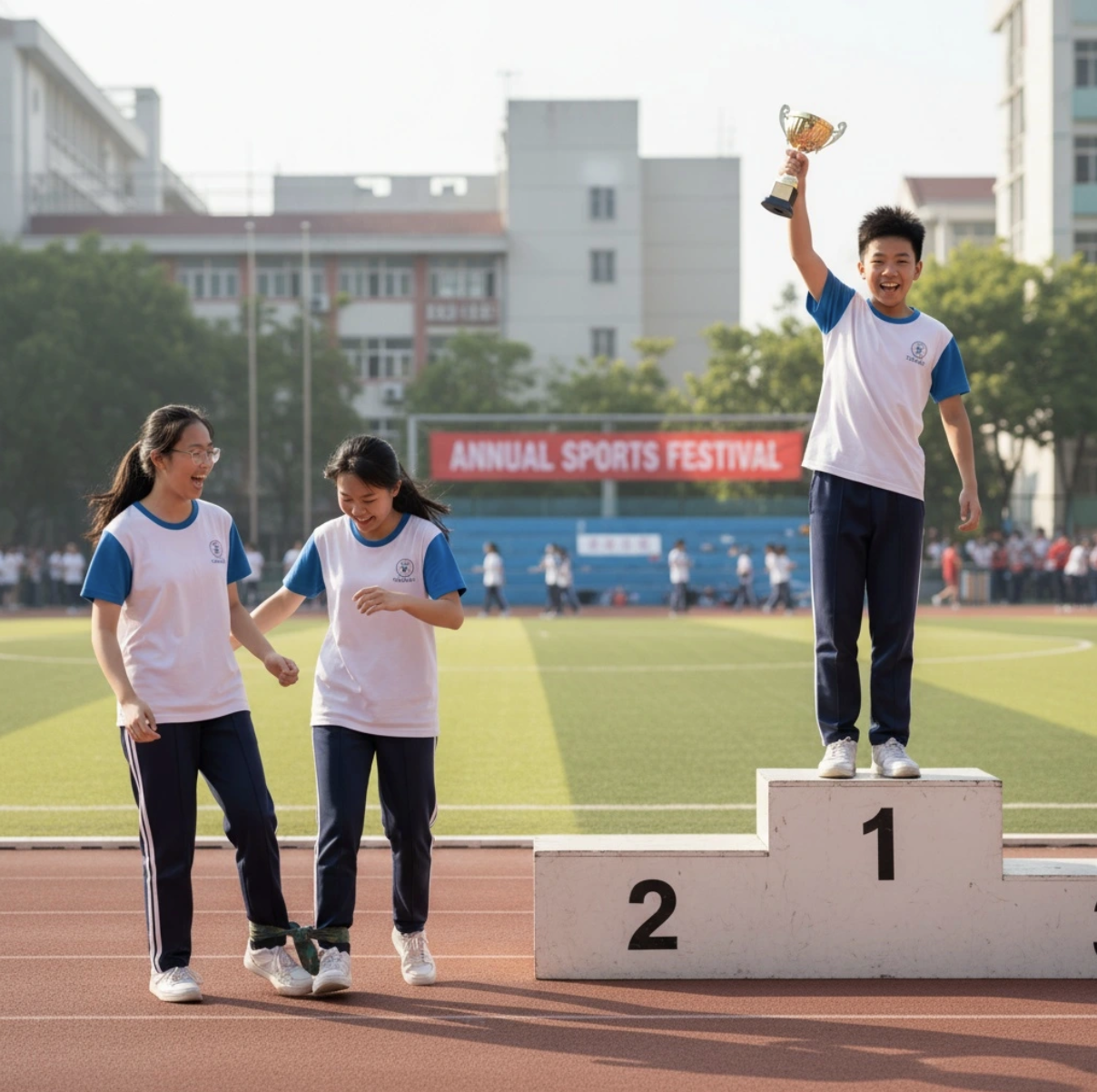}
  \captionof{figure}{P9 in a sports meet with high school classmates.}
  \label{fig:5}
\end{minipage}\hfill
\begin{minipage}[t]{0.24\textwidth}
  \centering
  \includegraphics[width=\linewidth]{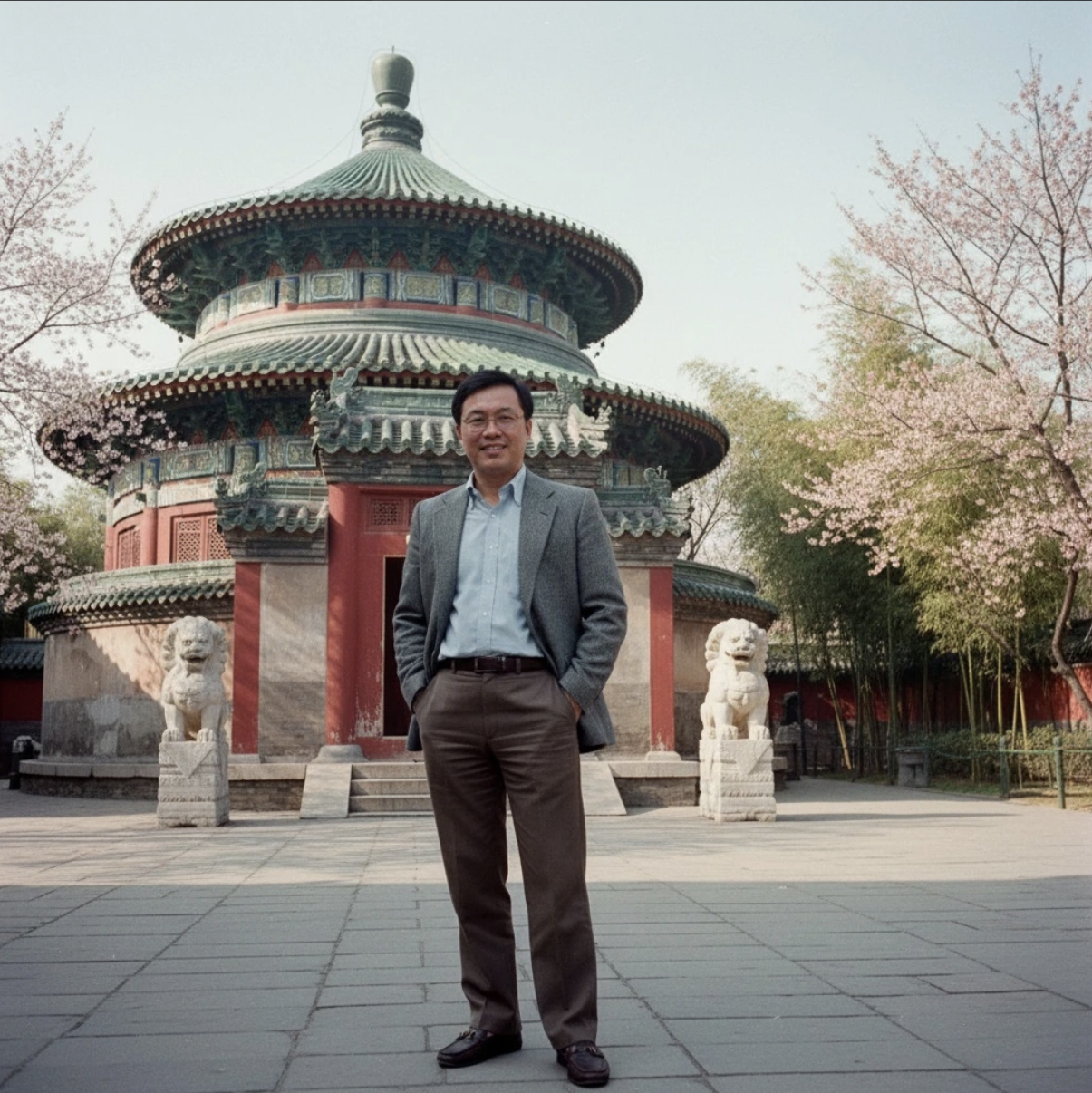}
  \captionof{figure}{P17 attempted to recreate a photo taken with a tourist attraction.}
  \label{fig:6}
\end{minipage}\hfill
\begin{minipage}[t]{0.24\textwidth}
  \centering
  \includegraphics[width=\linewidth]{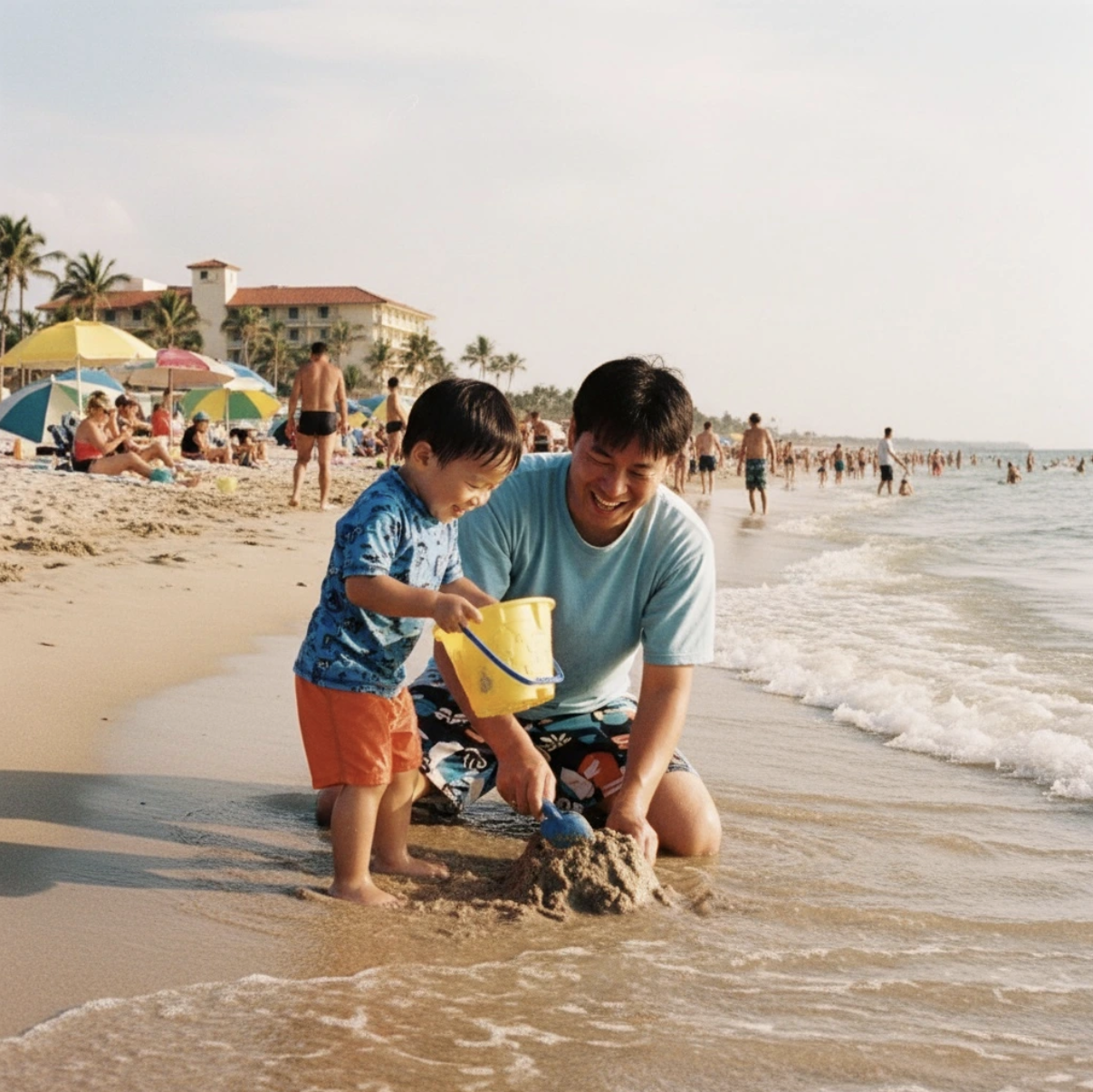}
  \captionof{figure}{P15 and his father played together by the sea.}
  \label{fig:7}
\end{minipage}\hfill
\begin{minipage}[t]{0.24\textwidth}
  \centering
  \includegraphics[width=\linewidth]{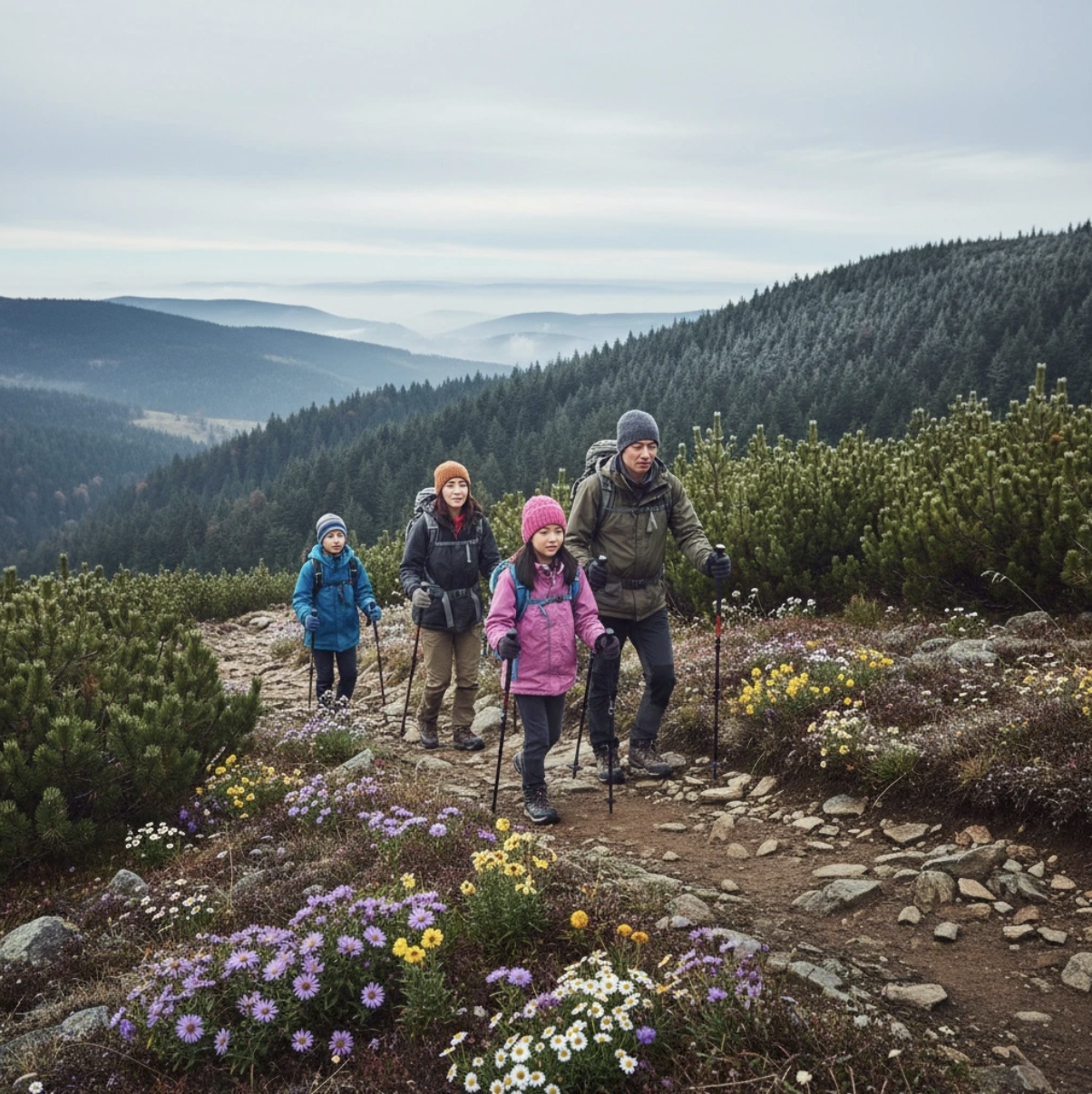}
  \captionof{figure}{Image of P8 hiking with family members, after multiple redraws by P8.}
  \label{fig:8}
\end{minipage}

\end{figure*}

The uncontrolled, unexpected elements that occasionally appear in images became intriguing aspects for participants. They found these \textit{``unexpected elements''} sometimes frustrating, yet other times pleasantly surprising. P2 attempted to recreate her first experience playing video games as a child  (Figure~\ref{fig:p2-games}). The retro computer in the image reminded her of when her parents first brought a computer home, prompting further recollections of what the computer desk looked like at the time. This brought the generated image closer to her memory. In other cases, these unexpected elements sparked reflections extending beyond the memory itself. P9 attempted to recreate a memory of participating in a sports meet with high school classmates  (Figure~\ref{fig:5}), but the generated image triggered associations like \textit{``I was terrified of public speaking as a child... I was so shy back then...''} In this case, these uncontrollable visual elements prompted the participant to reflect on and evaluate their own life. P17 attempted to compensate for a long-standing regret of not having a photo taken at a tourist attraction years ago (Figure~\ref{fig:6}). The architecture depicted in the generated image reminded him of some \textit{``ancient legends,''} prompting P17 to engage in additional discussions about these stories with the experiment organizers. Although these reflections sometimes had no direct connection to the memory itself, they prompted participants to engage in deeper contemplation.

\subsection{The Impact of Tangible Interface}
\label{subsec:impact-tangible-interface}
In addition to the positive effects of the generated images, the Memory Printer offers participants a novel human–AI interaction experience through its tangible form, providing them with more structured and controllable ways to engage with generative AI. 

\subsubsection{Physical Form and Material Appeal}
Participants reported that the physical presence of the Memory Printer created a unique appeal distinct from web-based AI tools. The device actively attracted them to engage with it. For example, P12 highlighted the aesthetic value of the Memory Printer, suggesting that: ``\textit{If the prototype were placed in my home, just seeing it would subconsciously make me wonder whether there was a memory I wanted to re-create.}'', and that “\textit{It could serve as a decorative object in the living environment}''. P7 emphasized the entertainment aspect of the device: ``\textit{Perhaps using AI to reconstruct memory scenarios could become an enjoyable daily activity. When I feel bored, I might use it as a form of leisure.}'' For some participants, the physical presence of the device imbued the act of remembering with a sense of ritual, distinguishing it from casual recollection or routine image generation. P15, for instance, attempted to reconstruct a childhood memory of playing with his father by the seaside (Figure~\ref{fig:7}). Although he often recalled this memory, he rarely considered specific details—such as what exactly happened that day or what his father was wearing. The tangible, deliberate nature of interacting with the Memory Printer led P15 to frame this remembering as ''\textit{making amends for the past,}'' prompting him to organize his narrative more thoughtfully than he would with a conventional digital tool. 

The material choices also received positive feedback. P19 noted that its wooden casing reminded her of the animation ``\textit{The Flintstones}'', where characters live in the Stone Age yet possess modern technologies ``\textit{packaged}'' in primitive forms, such as wooden cars or stone cell phones. She found this contrast highly amusing. Participants also reported that the wooden structure conveyed warmth and a pleasant tactile quality. As P9 observed: ``\textit{I like the smooth texture of birch wood. It does not feel as rigid as metal or plastic.}''
 
\subsubsection{The Role of Scraper in Human-AI Interaction}
None of the participants had prior hands-on experience with screen printing, though most had encountered it through videos or had long been curious about it. This familiarity without direct experience created an ideal starting point: participants approached the Memory Printer with curiosity rather than apprehension, eager to explore the interaction. Once engaged, they found that the screen printing metaphor made GAI interaction more intuitive and accessible. By incorporating alternative interaction paradigms grounded in familiar physical actions, the Memory Printer lowered the entry barriers for engaging with generative AI as an interactive, iterative process, and enhanced users' sense of control. P12 noted: \textit{``What has always prevented me from understanding generative AI is the complexity of the concepts... while I found the silk print easier to understand...''} P4 commented: \textit{``This prototype gives me a sense of a completely different possibility from the generative AI tools I have used before.''}

During the process, participants enjoyed interacting with it, repeatedly pulling the scraper to make the image disappear and reappear, and comparing the original with the regenerated version. The scraper transformed a simple task into an engaging activity, while also encouraging participants to observe the final images more carefully. This hands-on control created a stronger sense of agency than clicking buttons or typing prompts---participants felt they were actively shaping the output through body movement.

\begin{figure*}[t]
\centering
\begin{minipage}[t]{0.24\textwidth}
  \centering
  \includegraphics[width=\linewidth]{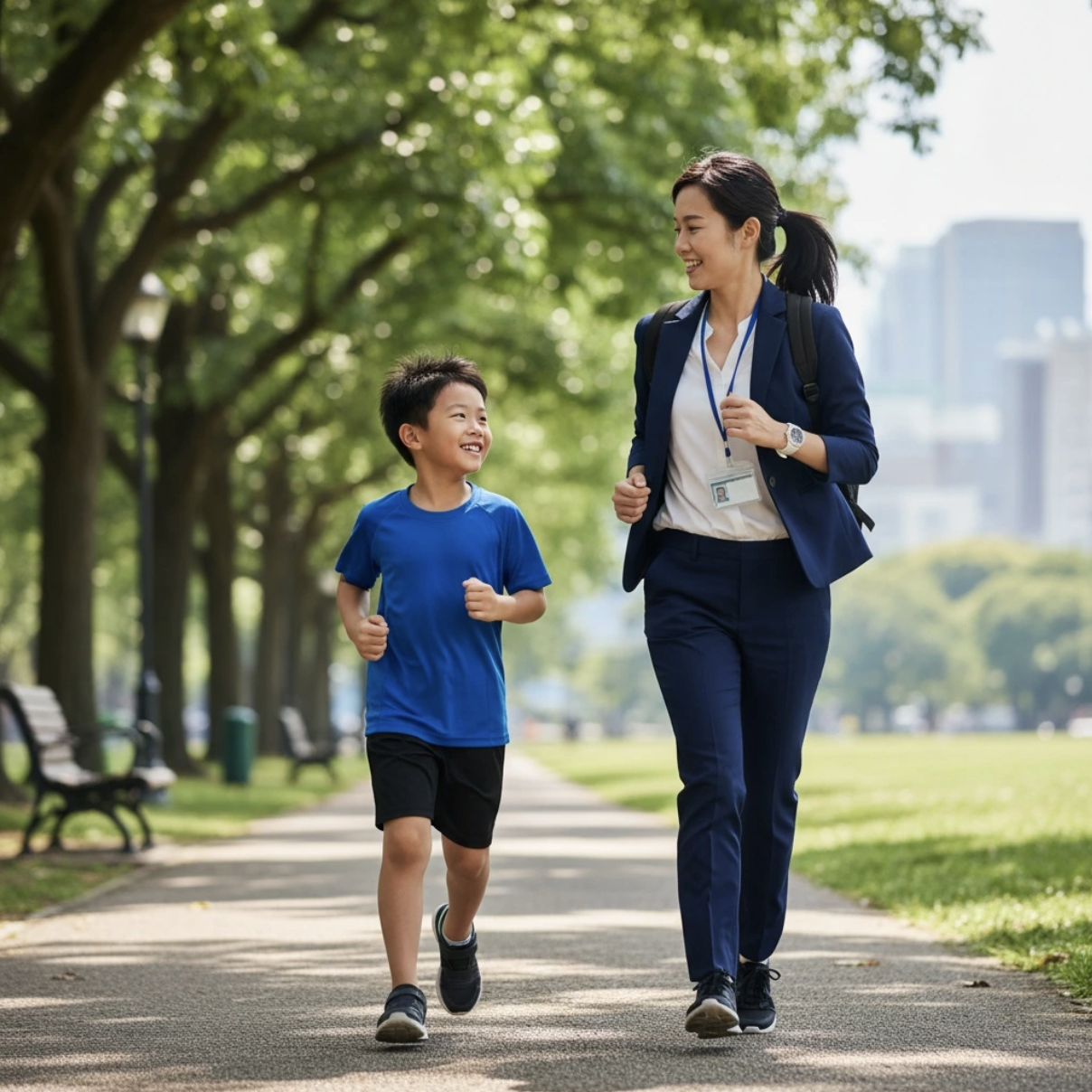}
  \captionof{figure}{The scene of P14 running with his mother, where the female figure in the image underwent multiple revisions by P14.}
  \label{fig:9}
\end{minipage}\hfill
\begin{minipage}[t]{0.74\textwidth}
  \centering
  \includegraphics[width=\linewidth]{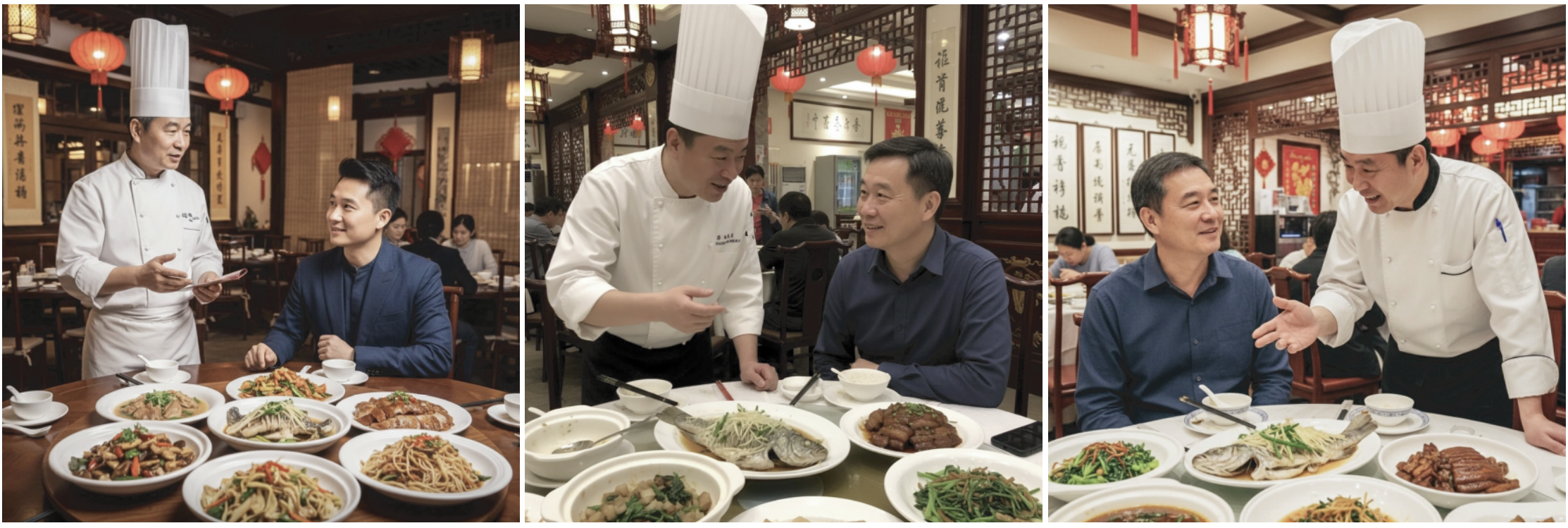}
  \captionof{figure}{Participants repeatedly attempted to generate images, but the element ``red lantern'' consistently appeared in the images.}
  \label{fig:10}
\end{minipage}
\end{figure*}

The introduction of the scraper deliberately slowed down the image generation process, and this enforced slowness proved valuable for reflection. P14 attempted to reconstruct a memory of running with his mother during her work breaks (Figure~\ref{fig:9}). As he moved the scraper to compare the modified and unmodified depictions of his mother, he reflected aloud: ``\textit{I didn't understand why she was running in her work clothes… Later, I realized it was because she didn't have enough time to change.}'' These reflections subsequently evolved into an acknowledgment of his mother's hard work and his own lack of time spent accompanying her. P14 emphasized that the slowness brought by the scraper ``\textit{gave me the opportunity to observe the modifications carefully,}'' without this slow, deliberate movement, he “\textit{would not have voiced such reflections.'' } In this case, the scraper's motion prompted P14 to move beyond recalling the event itself toward expressing a deeper emotional appreciation for his mother---a shift from factual recollection to meaningful reflection. 

We further observed that the scraper also showed potential in terms of harmful content moderation. Participants who were informed about possible disturbing outputs intentionally slowed their scraper movements, giving themselves more time to prepare. P6 explained that she deliberately paced her interaction, which ``\textit{provided more time to prepare for potentially harmful content.}'' Some participants questioned whether this approach might keep users in a state of constant anxiety—an issue that requires further investigation. Nonetheless, within the scope of the study, the scraper heightened participants’ awareness and prevented generated content from being presented to them too abruptly. However, since no harmful content appeared during the test, additional research is needed.

Overall, introducing the scraper as a component for image control enabled participants to move beyond purely screen-based interactions, encouraging more natural bodily movements that supported them in entering a reflective state of remembering. The scraper also provided a window for participants to carefully compare image details, allowing them to notice subtle aspects that might otherwise go unnoticed. From a usability perspective, the scraper transformed the ``\textit{emotionally flat and disengaging}'' process of AI image generation into an intuitive and enjoyable ''screen-printing'' experience, making GAI tools more approachable for participants while giving them greater control over the images.

\subsubsection{Layered Redrawing Supports Detailed Reflection}
The layered redrawing function provided participants with a structured approach to modifying AI-generated images. Instead of simply clicking buttons to regenerate entire images, users were required to identify, redraw, and describe specific areas they wanted to modify in different layers, while controlling the image update through the movement of the scraper. This method proved particularly useful for complex image modifications where multiple elements needed adjustment.

For example, P8 shared a memory of hiking with family members, in which he vividly described the day's weather and the surrounding environment. The image generated by the AI, however, contained contained numerous elements and diverged significantly from P8's perception. During the modification process, the layered modification paradigm enabled P8 to decompose the image into foreground, middle ground, and background, which in turn provided a clear strategy for making changes and prevented aimless adjustments. Rather than feeling overwhelmed by the mismatched image, he had concrete tools to systematically address specific elements, resulting in a stronger sense of control (Figure~\ref{fig:8}).

Feedback indicated that this method supported deeper visual inspection. By breaking down the modification task into manageable layers, participants could focus their attention on one aspect of the image at a time, leading to more thoughtful and deliberate adjustments. The layered modification function thus not only enhanced sense of control but also scaffolded a more structured cognitive process for memory reconstruction, helping participants organize their recollections in a spatial and systematic manner.
 
\subsubsection{Slow Design as a Means to Support Reminiscing}
Guided by slow design, we incorporated both the scraper-based interaction paradigm and the concept of layered modification into the Memory Printer, with the aim of prompting users to engage in more detailed observation and reflection. Rather than enabling rapid, click-based iteration typical of web-based GAI tools, these design elements deliberately slowed down the generation process, creating temporal and cognitive space for deeper engagement with memories. While the scraper controlled the pace of interaction, the layering system encouraged users to think systematically about what to modify and in what order. Taking P8's case as an example, the layered approach prompted a methodical examination of the scene. As he moved the scraper across the image to compare different versions, P8 articulated observations he might not have otherwise considered: \textit{``Although the mountains are cold at this time of year, the grass should still be very green...I like the pine trees in the distance, but they should be larger...''} The combination of slow physical movement (scraper) and structured modification (layering) created a rhythm of interaction that supported both redraw precision and emotional depth.

Together, these elements transformed memory reconstruction into a deliberate, contemplative practice. Participants reported that the slower pace, rather than being frustrating, enhanced their enjoyment and engagement. P22 noted that using the Memory Printer made him feel like he was \textit{``creating artworks about the past days,''} a characterization that reflects how slow design reframed reminiscence from passive recall into active, meaningful creation. The ``slowness'' gave participants permission to linger with their memories, to notice details they had forgotten, and to reflect on the emotional significance of past experiences. This stands in contrast to the rapid iteration cycles of most GAI tools, where the focus often remains on technical output quality and efficiency.

From a design perspective, in our work slow design principles proved particularly well-suited to reminiscence activities, where depth of reflection matters more than speed of output. By deliberately incorporating temporal friction through the scraper and cognitive structure through layering, the Memory Printer created conditions for what might be called ``slow remembering'', a mode of engagement that prioritizes contemplation, emotional connection, and meaning-making over efficiency.

\subsubsection{Physical Printing as Meaningful Souvenir}
The printing function transformed the digital interaction into a tangible outcome independent of the digital system. P21 likened this process to waiting for a Polaroid photograph to develop: \textit{``Even though I already know what the final image looks like, the waiting process itself is still enjoyable...it gives me more time to talk about what just happened.''} This deliberate delay, built into the manual printing mechanism, created a final space for reflection, allowing users to revisit the entire interaction and contemplate how they gradually reconstructed the memory. Unlike instant digital saving, which often feels perfunctory, the physical act of lifting the device and waiting for the print to emerge imbued the conclusion of the activity with a sense of ceremony. The waiting period became an opportunity for participants to narrate their memories further to researchers or companions, extending the reminiscence activity beyond the moment of image finalization.

Feedback also suggested that this waiting period not only facilitated further conversation but also imbued the outcome with added value. Participants described the printed result as both a \textit{``surprise''} (P8) and a \textit{``gift.''} (P11) They perceived that they were receiving more than just an experiential journey of recalling memories. This interpretation resonated particularly strongly with participants who saw the Memory Printer as a way to compensate for the absence of photographs from certain past experiences, filling gaps left in their personal histories. The physical print gave material form to memories that had previously existed only in mental or digital space, creating what participants described as \textit{``photos from the past''} (P17, P15) that represented both the memory itself and the process of its reconstruction.

\subsection{Challenges and Concerns}
\label{subsec:challenges-concerns}

\subsubsection{Concerns About AI-Induced False Memory Formation}
The majority of participants expressed concerns that AI-generated images might interfere with or overwrite their original memories. P14 articulated this anxiety: ``\textit{(AI-generated images) might merge with my original memories... eventually overwriting what I actually remembered.''} P11 used the term ``\textit{memory contamination}'' to describe this risk, noting: ``\textit{I think we need to remain constantly aware of this and maintain a clear-headed attitude... to distinguish between the two.''} However, some participants held different opinions. P16 offered a different view: ``\textit{This is a complex issue. I know this (AI-generated image) is definitely fake, but the emotions it evokes are real}.'' He argued that there is no such thing as ``\textit{100\% accurate memory}''---memories inevitably distort over time, and AI simply provides one pathway for this ``\textit{evolution.}'' This perspective reframes AI image intervention as a natural extension of memory's inherent malleability rather than ``\textit{memory contamination.}'' Several participants proposed possible interventions to mitigate false memory risks. P13 suggested that photorealistic rendering might exacerbate the problem: ``\textit{If the elements in the image were rendered in a 2.5D style, it could preserve the atmospheric quality I appreciate while avoiding (the false memory problem).}'' While photorealistic AI-generated images can provide users with more memory cues and trigger associations, their high fidelity also increases the risk of ``\textit{memory contamination.}'' 

\subsubsection{Discomfort with Unfamiliar Visual Representations} 
When depicting humans, AI-generated characters often appeared as unfamiliar representations, which many participants found unsettling yet unavoidable. P2 commented: ``\textit{Although the furniture and equipment in the image are close to my memory, the people are completely different...like myself and my family in a parallel universe}.'' Across participants, repeated attempts to refine human appearances frequently failed to produce satisfactory results, leading to frustration and a diminished sense of immersion. Some participants chose to abandon further refinement of human figures, while others expressed disappointment when their efforts did not yield recognizable likenesses. In response, participants commonly avoided incorporating detailed portraits of people unless the generated figures bore at least a partial resemblance to themselves or their family members. By contrast, generating representations of pets or inanimate objects was generally perceived as more feasible. For example, P13 was able to obtain a satisfying image of her dog by describing its breed, an experience she described as particularly delightful. 

While 23 participants accepted this limitation and focused on atmospheric experiences, one participant (P20) found it unacceptable. P20 attempted to use AI to recreate a scene of playing with their grandfather, but the generated figure bore little resemblance to his grandfather. He believed, ``\textit{this is disrespectful to my grandfather}.'' This discomfort stemmed from their intimate relationship with their grandfather, making it difficult to accept ``\textit{a stranger replacing (my) grandfather}.''

\subsubsection{AI Injustice and Stereotype} 
Stereotypical content frequently appeared in AI-generated images, particularly for specific cultural contexts. P13, whose father is a chef at a Chinese restaurant, provided a representative case. In the first phase of testing, P13 attempted to recreate his experience of visiting his father's workplace for the first time. In the second phase, P13 attempted to recreate a scene of his father explaining a dish to customers, which took place in the same Chinese restaurant. Despite never mentioning `red lanterns' in his descriptions, both AI models (Krea1 model in the first phase and Flux model in the second phase) repeatedly inserted this element into generated images (Figure 18).  In the first phase, P13 complained: ``\textit{I explicitly said kitchen... how could there be lanterns in a kitchen...}'' and stated that the red lanterns ``\textit{made the whole scene very unrealistic... like catering to some stereotype...}'' In the second phase, the repeated occurrence of similar issues made P13 feel this problem was ``\textit{very time-consuming... frustrating...}'' Although P13 ultimately completed the user test, he described the process as ``\textit{struggling against these stereotypes}'' rather than, like other participants, describing it as a pleasant and interesting creative activity.

Although the layered redrawing function provided P13 with tools for layered modifications to some extent, these elements still brought undeniable negative experiences to P13's experience. This typical case reveals issues that transcend the interface itself. Despite Memory Printer's deliberate design of mechanisms to help users control images and restore a sense of agency, participants still encountered persistent cultural stereotypes embedded in the underlying logic of the AI model. Memory Printer could not overcome these limitations. The AI formed a difficult-to-eliminate association between ``Chinese restaurant'' and ``red lanterns,'' which became a negative factor in P13's case. When P13 stated that lanterns were ``\textit{frustrating}'', this feedback demonstrated how algorithmic bias can undermine the goal of ``supporting memory reconstruction'': the AI system no longer scaffolds recollection but instead imposes external narratives, bringing negative experiences.

\subsubsection{Privacy and Data Security Concerns}
Regarding the issue of AI generating human images from photos, some participants inquired whether they could use their own photos to train the AI and correct this problem. This raised another concern shared by most participants: Could the AI misuse their data and violate their privacy? When we presented this idea to participants during interviews, 20 out of 24 participants expressed reluctance to provide their own or their family members' photos for AI training. They feared potential misuse of the photos, and the risks posed by deep-fake technology made them more inclined to avoid having their likeness appear in AI-generated images. Among the remaining 4 participants, one considered privacy breaches inevitable, while the other three, though finding the concept intriguing, advocated for stricter AI limitations—such as granting only specific access permissions. 

\section{Discussion}
Our research indicates that the memory printer has yielded positive results in enhancing reminiscence. Participants were not only able to partially reconstruct memory fragments lacking photographs using GAI, but the memory printer also facilitated deeper recollection and reflection. The tangible interface played a crucial role. Designed specifically for memory characteristics, the Memory Printer supported recollection through its slow-paced interaction, while also providing participants with an easy-to-learn, engaging interactive experience. During interviews, some participants expressed concerns about AI intervening in the recollection process, particularly regarding the authenticity of AI-generated images and data security issues. This section further discusses the research findings, summarizes the strengths and limitations of generative AI in facilitating recollection, and outlines future research directions.

\subsection{Using GAI to Facilitate Self-Expression}
The modified AI images also served as a medium for conveying emotions, helping participants express feelings that were difficult to articulate and providing them with a way for self-expression. Some participants felt this interaction closely resembled artistic creation. GAI assisted participants in transforming abstract inner emotions into concrete images, which were then gradually refined through interaction with the images. Ultimately, the images served as emotional carriers shared with the experiment organizers.
Participants also treated the generated images as mementos of their reminiscence. Based on this, they suggested adding supplementary recording features to the system---such as documenting the entire recollection process through various media (text, audio recordings)---to make the recollection process itself a keepsake, rather than just the generated image. In this process, AI-generated images become just one component of a broader commemorative practice. Users retain control over which media to preserve—positioning AI as a facilitator rather than the sole author of the memento.

\subsection{Designing with the Characteristics of AI}
Our research draws upon two core characteristics of AI: first, its ever-increasing capacity for content generation~\cite{Hartmann2025}; and second, the inherent uncontrollability of such generative processes~\cite{Hicks2024}. Although researchers are making advances in improving controllability~\cite{CavalcanteSiebert2023}, errors and the need for adjustments persist throughout this process. Rather than attempting to eliminate this uncontrollability, we intentionally embrace it as a design element. The unpredictability of generated content introduces a sense of surprise into the image modification process. When participants compare modified images with their own memories, they are naturally prompted to comment on the images, which in turn can spark new conversations. This phenomenon was evident for one older participant (P17), who expressed anticipation upon receiving the images. Given our limited age diversity, we do not draw age-related conclusions; future work should examine whether this pattern holds for older populations. The excitement stemmed not only from the desire for communication but also from curiosity about the content itself. We therefore recommend that designers ground their approaches in the inherent qualities of AI, exploring how to harness what might appear as ``undesirable'' traits and transform them into solutions.

\subsection{Human–AI Interaction Beyond the Screen}
In our user studies, participants demonstrated limited interest in conventional web-based AI tools, finding them not only difficult to learn despite simplification, but also unappealing. Moreover, the homogenization of human–AI interaction has constrained the integration of AI tools into diverse contexts. In our research, although we employed touchscreens and the traditional image generation process (i.e., inputting prompts and generating images), we deliberately minimized participants’ awareness of this underlying mechanism. Instead, we provided intuitive and tangible methods for image control, thereby enhancing participants’ sense of agency while creating space for reflection. Participants responded positively to this mode of interaction and expressed interest in seeing more such examples in the future.
We argue that combining AI tools with existing interaction paradigms presents a promising direction for research. This approach offers users opportunities for engaging with AI beyond the confines of the screen, thereby supporting both learning and trust-building in human–AI relationships. 

\subsection{Restoring Sense of Agency in Human-AI Interaction}
Our results demonstrate that Memory Printer's interaction modalities enable stronger engagement and control compared to web-based tools. Sense of Agency (SoA) refers to an individual's subjective experience of controlling their own actions and, through them, events in the external world. Prior research shows that it is influenced by multiple cues, including motor signals from movement production itself and external contextual cues that can trigger reflection—these two types of cues influence the generation of agency from internal and external perspectives, respectively~\cite{moore2012sense}. As described in Section~\ref{subsec:User Agency in Human-AI Interaction}, web-baed GAI tools require users to express intent through prompt engineering. This highly automated, unobservable, screen-confined interaction paradigm disrupts both components of the agency experience: control over one's own actions (executing meaningful, active interactions) and control over external outcomes (predicting generation results and attributing them to specific actions). Memory Printer proactively provides users with these multidimensional agency cues, compensating for the sense of agency disrupted by virtual interfaces based human-AI interaction paradigms. In this section, we analyze from these two perspectives how Memory Printer compensates for agency through design. 

\subsubsection{Enhancing Control Over Actions}
Memory Printer reintroduces embodied behavior into the generative AI workflow through physically grounded interaction modalities. Rather than confining users to text input and button clicks, the system requires users to actively move the scraper and describe their drawing areas during the reminiscence process. The introduction of the scraper is crucial for restoring agency. The process of moving the scraper creates tactile and kinesthetic engagement for users, converging proprioceptive, haptic, and visual modalities. Participants' feedback strongly supports this point. When using Memory Printer, multiple participants felt they were ``engaging in artistic creation'' and that ``the scraper made them feel like they were really doing printing work,'' whereas in the first phase, participants viewed their activities more as completing instructions from the experimental organizers rather than actively engaging in "artistic creation." Memory Printer helped participants achieve a role transformation from passive recipients to active creators, stimulating their Sense of Agency.
Beyond the diversity of physical actions, Memory Printer also grants users temporal autonomy. Most web-based GAI tools impose a binary temporal structure: users click ``generate,'' and results subsequently appear. By introducing the scraper, users can actively control the rhythm of image revelation through the scraper's movement, deciding when to pause for inspection, when to accelerate through satisfactory areas, and when to reverse direction for comparison. All participants spontaneously moved the scraper repeatedly during testing to control image display, attempting to compare modified regions and confirm image details. P8's thought process progressed incrementally with scraper movement: ``\textit{Let me pull back to check... yes, now move forward... the mountains look greener... now let's modify the size of the trees...}'' This temporal control transforms passive reception of images into active investigation. Rather than accepting or rejecting a complete, instantaneously presented result, users progressively build their understanding through deliberate, self-paced revelation. This enhances users' Sense of Agency from another dimension.

\subsubsection{Enhancing Control Over External Outcomes}
Memory Printer enhances users' sense of control from the perspective of external outcomes primarily through three interrelated mechanisms. First, the scraper's movement transforms instantaneous image appearance into a continuous, real-time process. Second, the layered redrawing function gives image reconstruction a sense of spatial organization and order. Users do not need to reconstruct the entire scene each time during image generation but only redraw the regions they feel need modification. As P14 stated: ``\textit{I can see exactly what each change does... if I don't like the new element, I know it was caused by layer 3.}'' Participants spontaneously adopted `foreground-middleground-background' terminology, indicating conscious organization of their redrawing activities. Finally, the reduction in cognitive load facilitates deeper image construction. Unlike in the first phase, where participants needed to regenerate the entire image each time and make substantial modifications to prompts, layered redrawing allows them to focus on only one detail at a time, reducing the cognitive load of individual image modifications. The layered redrawing function also enables participants to fully attend to single modification elements, making it easier to perceive and remember the connection between specific prompts and modification results. This enhanced causal tracking strengthens outcome-level agency cues.
Memory Printer's image printing function extends participants' control over external outcomes beyond the screen. Prior research suggests that physical artifacts bring special experiences that transcend digital files \cite{ishii1997tangible}. Participants viewed printed generated images as ``compensation for memories without photos'' and ``souvenirs of created memory scenes.'' They could ultimately decide whether to preserve them and in what medium. This transformation of image display medium brought participants a stronger Sense of Agency.

In this work, although the AI model's operating principles and specific training datasets remain black boxes to participants, Memory Printer provides users with compensatory Sense of Agency from two dimensions and achieved positive user feedback. This demonstrates that in human-AI interaction, designers can still help users restore agency through carefully designed interaction methods.

\subsection{Navigating Risks in GAI-Enhanced Reminiscence}
While our findings demonstrate the potential of using generative AI for reminiscence support, they also surface critical concerns that demand careful consideration. The emotionally sensitive nature of memory work amplifies risks that might be tolerable in other GAI applications. This section synthesizes these concerns and discusses their implications for future design and research.

\subsubsection{The False Memory Dilemma}
The potential for AI-generated images to contaminate authentic memories represents perhaps the most fundamental challenge. Prior work shows that AI-edited images can significantly increase false memory incidence \cite{Pataranutaporn2025}, and photorealistic quality simultaneously enables valuable memory cues while increasing integration risk.

This creates an inherent tension: higher visual fidelity provides richer memory cues but simultaneously blurs the boundary between generated and authentic content, potentially facilitating what participants termed ``memory contamination.''

As P16 noted, emotions evoked by AI-generated images can be genuine even when visual content is not. Rather than attempting to preserve some notion of memory purity, future designs might focus on helping users maintain a clear understanding of which elements derive from AI generation versus personal recollection. Several design strategies merit exploration: such as visual markers that distinguish AI-generated content without disrupting emotional engagement, or interaction patterns that reinforce the constructive nature of the reminiscence process. Longitudinal research is needed to understand how repeated exposure to AI-generated memory images affects autobiographical memory over time, and whether design interventions can effectively mitigate this situation.

\subsubsection{Algorithmic Bias as Narrative Imposition}
Our findings reveal how algorithmic bias transforms from a technical problem into an experiential one. P13's encounter with persistent `red lantern' stereotypes illustrates this: rather than scaffolding recollection, the AI imposed external narratives on his personal history. For P13, each AI-imposed lantern represented not just a technical failure but a small act of cultural erasure. 
We trace this issue to the provenance and composition of training datasets for contemporary generative models. When these models encounter culturally-marked keywords (such as Chinese restaurant), they retrieve patterns from statistically dominant representations, in the current case, visual and touristicized representations of Chinese restaurants, rather than the mundane, lived realities users seek to reconstruct. For supporting reminiscence, this is particularly dangerous: the system does not merely fail to understand diverse contexts; it actively overwrites diverse contexts with hegemonic visual symbols.
These issues clearly contradict human-centered AI principles and serve as a warning: AI unfairness and flawed decision-making can negatively impact users who trust AI, ultimately eroding their confidence and potentially leading to AI rejection. Interventions such as Memory Printer's layered redrawing can provide users with tools to resist algorithmic impositions, but they cannot address the root cause. More fundamental solutions require attention to training data composition, model fine-tuning for cultural diversity, and potentially user-specific adaptation. Designers working in this space should also consider how to communicate these limitations transparently, helping users understand when and why AI outputs may diverge from their intentions due to systemic biases rather than individual prompt failures.

\subsubsection{Privacy, Trust, and the Boundaries of AI-Mediated Communication}
Participants' reluctance to provide personal photographs for AI training—even when it might improve character representation—reveals a boundary condition for GAI-enhanced reminiscence. The intimacy of memory creates heightened sensitivity to data practices acceptable in other contexts. Participants weighed the potential benefits of more accurate visual representations against risks of data misuse, and the majority chose to preserve privacy at the cost of immersion.

This finding suggests that trust in GAI systems for reminiscence cannot be established through technical capability alone. Participants expressed concerns about deep-fake misuse, unauthorized training, and lack of transparency in how their data might be used. These concerns impose what we term ``invisible constraints'' on human-AI interaction: users self-limit their engagement with AI systems based on perceived risks, even when more open engagement might yield better outcomes.

Addressing these constraints requires attention to both technical and communicative dimensions. Local deployment and on-device processing might can reduce data exposure risks, while transparent explanation of data practices might can help users make informed decisions about their level of engagement. However, our findings suggest that for many users, no level of technical assurance may be sufficient to overcome reluctance to share intimate visual data. Future designs might explore alternative approaches that achieve personalization without requiring personal data, for example, using verbal descriptions, sketches, or authorized reference images to guide image generation.

\subsubsection{Toward Responsible Design of GAI for Emotionally Sensitive Contexts}
Synthesizing across these concerns, we identify several principles for responsible design of GAI tools in emotionally sensitive contexts such as reminiscence:

First, Designers must be aware of the potential consequences of interacting with AI. In efficiency-focused GAI applications, inaccurate outputs primarily waste time; in emotionally sensitive applications, they can cause psychological harm, reinforce stereotypes, or contaminate precious memories. Designers need to recognize that Human-AI interactions can have varying degrees of impact in different context and respond consciously to these impacts through design.

Second, transparency must extend beyond technical explainability to include honest communication about system limitations. Users engaging in memory work deserve to understand not only how the AI generates images but also what biases it may carry, what risks repeated exposure might pose for memory accuracy, and what happens to their data. Such transparency may reduce engagement in the short term but builds the foundation for sustainable trust.

Third, designers need to consider how to enhance user agency in the Human-AI interaction process through design, particularly in contexts where generated content may carry personal or emotional significance Our findings suggest that in the context of using GAI to support reminiscence, restoring user agency through multiple complementary mechanisms, such as tangible interaction, temporal control, and layered modification, can amplify the benefits that GAI offers. When participants felt they were actively shaping rather than passively receiving AI outputs, they engaged more deeply with both the tool and their memories. While further research is needed to examine whether these insights generalize to other domains, the rapid advancement of AI capabilities and the expanding scope of its deployment suggest that understanding how to design for agency in human-AI interaction represents a meaningful and increasingly urgent research direction.

These principles do not resolve the fundamental tensions inherent in applying GAI to emotionally sensitive domains, but they provide some references for navigating these tensions thoughtfully. As generative AI capabilities continue to advance, the HCI community bears responsibility for ensuring that these powerful tools are deployed in ways that genuinely serve human flourishing rather than merely demonstrating technical possibilities.

\subsection{Limitations and Future Work}
Our work has several limitations that provide directions for future research. First, our participant sample primarily comprised young and middle-aged adults (23 participants aged 22-31), with only one participant aged 57. While our design arguments about slowness, embodiment, and decomposed control are not age-specific, we acknowledge that age-related differences in technology familiarity, memory practices, and emotional connections to physical artifacts may influence how different populations experience Memory Printer. Prior research indicates that older adults often have stronger emotional connections with physical memory artifacts and tangible interaction paradigms, which may generate different responses to our design arguments compared to younger users. Future research should examine whether our design arguments hold across more diverse age groups, including older adults who are often the focus of reminiscence therapy research.


Second, our participants were drawn from two countries (China and the Netherlands) and did not include participants from other cultural backgrounds, which limits cultural diversity to some extent. Although participants from these two cultural backgrounds provided similar feedback in many aspects, they exhibited significant differences, particularly in areas such as AI injustice and stereotypes in AI-generated content.

Finally, our research revealed several critical concerns about using generative AI for memory scene reconstruction---the potential harm of false memories, the negative impact of algorithmic bias, and personal data privacy---that we identified but not not deeply investigate. Since this work focused on exploring design arguments about tangible, slow interaction with GAI in supporting everyday reminiscence and understanding the potential role of diversified human-AI interaction in this process, these ethical and technical challenges remain important directions for future research.

To address these limitations, we are planning to advance several initiatives. First, we will establish connections with local elderly care facilities to recruit older adult participants with varying levels of technical experience to support subsequent age-related research. We will also expand our recruitment channels to include participants from more diverse cultural backgrounds to understand how preferences vary across different regions in multicultural contexts. Finally, through human intervention measures, including local AI deployment and training custom AI models, we hope to reduce the occurrence of biased AI content and provide protection for personal data privacy, addressing these issues from a technical perspective. While the false memory problem is difficult to completely eliminate due to the nature of AI, in subsequent user testing, we will attempt to make participants aware of this issue through proactive notification and real-time reminders, and understand their perspectives on this problem.

\section{Conclusion}
This paper introduces the Memory Printer, a tangible design that combines traditional screen printing with generative AI to support reminiscing in everyday life. Adopting a Research through Design approach, we use the Memory Printer as a design exemplar to explore both the potential and pitfalls of GAI in supporting everyday reminiscence. Through an empirical study with 24 participants, we demonstrate that the Memory Printer can facilitate memory reconstruction and emotional expression, while also surfacing important concerns about false memories, algorithmic bias, and data security. 

From a sense of agency perspective, we analyze how tangible, slow, and layered interaction design can restore user control in human-AI interaction. Our findings provide empirical evidence that these design qualities work synergistically: the physical scraper enables embodied engagement and temporal autonomy, layered redrawing supports causal attribution and systematic modification, and the deliberate pace creates a reflective space that rapid web-based tools cannot provide. We then offer actionable insights for future design work in emotionally sensitive AI applications. From a methodological perspective, the knowledge generated through this process is not a universal design guideline but a grounded, experiential understanding of how slowness, tangibility, and layered control shape user experience. This exemplifies how RtD can contribute to HCI's understanding of emerging technologies by making abstract concepts tangible and debatable. Furthermore, we position the Memory Printer as a design exemplar that challenges mainstream assumptions in the development of generative AI. As these technologies proliferate, we face the risk of converging on a single interaction model---one that overemphasizes speed and automation. By grounding human-AI interaction in slow design and material engagement, the Memory Printer points toward a future where generative AI serves not just productivity, but also contemplation, connection, and personal meaning-making.

\bibliographystyle{ACM-Reference-Format}
\bibliography{memory_printer_chi26}

\end{document}